\documentclass[review]{elsarticle}

\usepackage{lineno,hyperref}

\usepackage[utf8]{inputenc} 
\usepackage[T1]{fontenc}
\usepackage{hyperref}       
\usepackage{url}          

\usepackage{latexsym}
\usepackage{graphics}
\usepackage{graphicx}
\usepackage{amsfonts,amssymb,amsmath}
\usepackage{dblfloatfix} 
\usepackage[export]{adjustbox}
\usepackage[ruled,vlined]{algorithm2e}
\usepackage[margin=1.0in]{geometry}
\usepackage{booktabs}      
\usepackage{nicefrac}       
\usepackage{microtype}     
\usepackage{caption}
\usepackage{subcaption}
\usepackage{catchfile}
\usepackage{totcount}
\usepackage{multirow}
\usepackage{array}

\newcolumntype{P}{>{\centering\arraybackslash}m{1.45cm}}
\newcolumntype{Q}{>{\centering\arraybackslash}m{0.90cm}}

\modulolinenumbers[5]

\begin{document}

\begin{frontmatter}

\title{Inferring Mobility of Care Travel Behavior From Transit Smart Fare Card Data}

\author[1]{Awad Abdelhalim\corref{mycorrespondingauthor}}\cortext[mycorrespondingauthor]{A. Abdelhalim and D. Shuman contributed equally to this work. A. Abdelhalim is the corresponding author.}
\ead{awadt@mit.edu}
\affiliation[1]{organization={Massachusetts Institute of Technology},
            addressline={Department of Urban Studies and Planning}, 
            city={Cambridge},
            postcode={02139}, 
            state={MA},
            country={USA}}

\author[2]{Daniela Shuman}
\affiliation[2]{organization={Harvard University},
            addressline={School of Engineering and Applied Sciences}, 
            city={Cambridge},
            postcode={02138}, 
            state={MA},
            country={USA}}

\author[1]{Anson F. Stewart}

\author[3]{Kayleigh B. Campbell}
\author[3]{Mira Patel}
\author[3]{Gabriel L. Pincus}
\affiliation[3]{organization={Washington Metropolitan Area Transit Authority},
            city={Washington},
            postcode={20024}, 
            state={DC},
            country={USA}}

\author[4]{Inés Sánchez de Madariaga}
\affiliation[4]{organization={Universidad Politécnica de Madrid},
            addressline= {Department of Urban and Regional Planning},
            city={Madrid},
            postcode={28040}, 
            country={Spain}}

\author[1]{Jinhua Zhao}

\begin{abstract}
Existing research underscores substantial gender-based variations in travel behavior on public transit. Studies have concluded that these differences are largely attributable to household responsibilities typically falling disproportionately on women, leading to women being more likely to utilize transit for purposes referred to by the umbrella concept of “Mobility of Care”. In contrast to past studies that have quantified the impact of gender using survey and qualitative data, we examine a novel data-driven workflow utilizing a combination of previously developed origin, destination, and transfer inference (ODX) based on individual transit fare card transactions, name-based gender inference, and geospatial analysis as a framework to identify \textit{mobility of care} trip making. We apply this framework to data from the Washington Metropolitan Area Transit Authority (WMATA). Analyzing data from millions of journeys conducted in the first quarter of 2019, the results of this study show that our proposed workflow can identify \textit{mobility of care} travel behavior, both in terms of (1) detecting times and places of interest where the share of women travelers in an equally-sampled subset (on basis of inferred gender) of transit users is 10\% - 15\% higher than that of men, and (2) finding women significantly more likely to exhibit a consistent accompaniment patterns with riders who are children, elderly, or people with disabilities. The workflow presented in this study provides a blueprint for combining transit origin-destination data, inferred customer demographics, and geospatial analyses enabling public transit agencies to assess, at the fare card level, the gendered impacts of different policy and operational decisions.\\
\end{abstract}


\end{frontmatter}

%\linenumbers

\newpage
\section{Introduction}
% Hook - why is this topic interesting
Research on women and gender in transport carried out in the last three decades suggests that women tend to interact with public transportation differently from other users \cite{de2013women}. Most of this difference in use can be attributed to the typical distribution of household responsibilities largely falling on women. Many of these studies are based on survey data and qualitative data, which show evidence that women spend more time and energy using public transportation and active mobility to conduct these typical gender-role-defined tasks \cite{mcgill_2022, montoya2020gender}, referred to by the umbrella concept \textit{“mobility of care”}. The concept was introduced by Ines Sanchez de Madariaga in 2009 \cite{de2016mobility} who further hypothesized in 2013 that the total number of such trips could be close to that of employment-related mobility \cite{de2016mobility}. The \textit{mobility of care} concept provides a framework for "recognizing, measuring, making visible, valuing and properly accounting for all the travel associated with those caring and home-related tasks needed for the reproduction of life”. The umbrella concept \textit{mobility of care} unveils trips related to the care of other persons and of the household that normally are made by women and hidden under other categories defining travel purposes, such as escorting, visiting, strolling, shopping, or other. Studies also demonstrate that women commonly make up the majority of riders on U.S. public transportation \cite{clark2017rides}. 

\begin{figure}[!h]
\centering
\includegraphics[width= .65\textwidth]{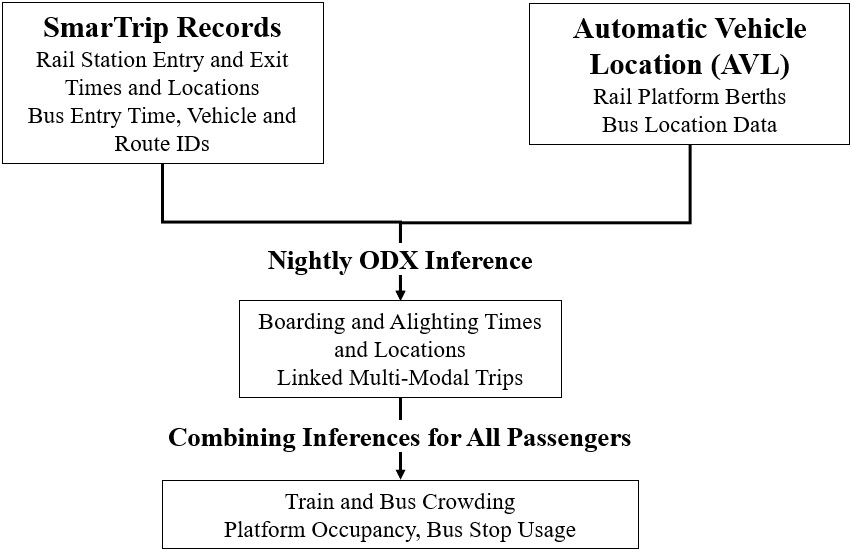}
\caption{\label{fig:odx} ODX deployment workflow in WMATA.}
\end{figure}

Recent advances in inferring passenger behavior from automated data collection systems provide a richness of information about how individuals pay and travel throughout the public transit system \cite{wilson2009potential, koutsopoulos_2017}. However, there is currently an evidence gap in evaluating the results of survey-based studies against those automated data collection sources and fare card data in particular. This research addresses that gap by making it possible to examine the extent of gender-specific travel behavior at the passenger-card level. While this gender-specific travel behavior has typically been studied through user surveys, which provide rich information about behavioral choices and why people are making the travel decisions they are, they can also be expensive to conduct on a large scale and take time to provide actionable evidence.  Whereas for automated fare card data, collection and analysis of substantially larger quantities of high-granularity passenger travel information can occur as early as the next day. A shift towards using big data, particularly disaggregate origin-destination data, can help validate existing hypotheses about the impact of gender on travel behavior and fare burden and reveal new findings as to how individuals, women particularly, actually use the system. Methods to infer passenger origin, destination, and transfer (ODX) patterns from combined automatic fare collection and automatic vehicle location data have been developed previously \cite{sanchez2017inference, gordon2018estimation}. The results of this disaggregate, comprehensive inference process provide for a detailed study of travel behavior within public transportation systems \cite{goulet2017measuring}. This work builds on a commercial ODX implementation, illustrated in Figure \ref{fig:odx}, built on the methods cited previously deployed at the Washington Metropolitan Area Transit Authority (WMATA).

Combining ODX data with demographic information is a powerful, highly granular way to observe and assess how specific groups of people move around a city using public transportation. For WMATA in particular, women are the majority: more trips are taken by women (58\% on Metrobus and 54\% on Metrorail), more unique customers are women (55\%), and more people in the region are women (51\%) (data from WMATA's customer surveys, and US census). Seemingly gender-neutral policies can have surprisingly different impacts, including unintended negative impacts on women. Gender disaggregated data, developed through this research, provides a new tool that WMATA can use to evaluate gender equity in fares and services. The first application, presented here, is to use this data to quantify the gender split of \textit{mobility of care} travel behavior. Based on the literature, we expect to observe more women conducting trips that qualify as \textit{mobility of care} trips. Developing a better understanding of \textit{mobility of care} travel, regardless of the gender of the caregiver, is paramount in informing agency decisions about stop, schedule, and infrastructure needs around the times and locations of these trip patterns that fulfill critical societal functions.

% Introduce the study at a high level

\subsection{Objectives and Contribution}
% Why is this specific study important
% What have we done differently from past studies
To the authors' knowledge, there are no major studies investigating \textit{mobility of care} travel behavior from ODX data, with studies in the field largely dependent on survey and qualitative data. The authors believe that a data-driven approach, based on the truly observed trip-making behavior complements the traditional approaches and provides public transportation agencies with highly granular analyses that can aid in guiding and evaluating operation and policy decisions. We propose and evaluate a workflow that successfully identifies users exhibiting \textit{mobility of care} travel behavior using WMATA's ODX data for a set of registered customer fare (SmarTrip) cards from the first quarter of 2019, providing a much-needed decision support layer to public transportation agencies.

\section{Related Work}
% Perhaps one paragraph on Washington DC specific literature
Studies have found women to have relatively distinct travel patterns compared to men. Women are more likely to travel on public transit generally, as well as off-peak (typically noon - 4 PM), for shorter distances, and in chained trips \cite{sanchez_zucchini_2019}. These findings on the characteristics of trips have been consistent across numerous studies, including a detailed 2019 study \cite{lametro2019} that validated many of these findings in the context of Los Angeles County. In a separate study, Gauvin et al. \cite{gauvin_gender_2020} used mobile phone data in Santiago, Chile and found that women tend to be more spatially localized than men, as well as more predictable in their travel patterns. Women were found to distribute their trips among a few highly preferred locations while men distribute their trips among many locations with almost equal probability. In other words, women can be found more frequently at their more visited locations. Mark and Heinrichs conducted a study in Buenos Aires concluding that gender-based trips typically follow separate pre-structured schedules, especially for women with young children \cite{mark_more_2019}. These findings were consistently observed in many cities across the world with a 2018 study that combined multiple surveys from 8 cities \cite{/content/paper/eaf64f94-en}. 

There are differences in travel patterns among women depending on employment status and household income status. Gauvin et al. found that the disparity in women's mobility patterns increases significantly when moving from the wealthiest to most deprived communities of Santiago \cite{gauvin_gender_2020}. Further, this study found that the gender mobility gap is significantly correlated with employment status, higher fertility rates, and larger households. Using travel diary data, Kwan (1999) found that employment status and how family obligations are divided play an important role in women's mobility patterns \cite{kwan_gender_1999}. Similarly, in a survey with approximately 25,000 households and 60,000 individuals conducted by the German Institute for Economic Research, it was found that the average trip length of part-time employed individuals, and especially homemakers, are lower than that of full-time employed individuals. However, when disaggregating these results by gender, women were found to have a 5\% increase in trips per day compared to men \cite{nobis_gender_2005}, demonstrating that the impact of employment on average trip length is higher among women. This phenomenon, however, does not apply to part-time working single women, who have fewer but longer trips than their men counterparts.

The majority of studies on gendered travel patterns rely heavily on survey data. Survey data has been a powerful tool to examine gender equity issues in systems. The International Transport Forum commissioned an aggregation of survey data from 8 different cities across the world in 2018 \cite{/content/paper/eaf64f94-en}. Using multinomial logistic regression analysis, this study explored the relationship between gender and transport mode choices. Some of the key findings were that women make more non-commute trips than men, their travel distances are shorter, and women are more likely to walk or cycle than men. 
The \textit{mobility of care} is a useful umbrella category that characterizes a pattern of trip-making. The first major study conducted on \textit{mobility of care} travel was made through a specifically designed survey conducted in the metropolitan area of Madrid \cite{sanchez_zucchini_2019}. To characterize the nature of \textit{mobility of care} trips, Sánchez de Madariaga and Zucchini designed and implemented a survey specifically conceived to be able to quantify \textit{mobility of care} trips in the metropolitan area of Madrid for the population aged 30-45. The main findings of this study demonstrate the original hypothesis posed by Sánchez de Madariaga in 2013: the number of trips under the umbrella notion of \textit{mobility of care} could be close to employment-related trips. The gender gap in care-related travel was much bigger than the gender gap in employment-related travel. A key aspect of this study was to properly measure care-related travel through question design. Care-related trips are often hidden under other headings including shopping, escorting, visiting, strolling, or others. This study specifically distinguished between forms of care-related travel.

Several transit agencies, including but not limited to Los Angeles Metro, Vienna, and Montreal, have used surveys to improve their services for groups of riders, including women. The Los Angeles County Metropolitan Transportation Authority (LA Metro) is a prominent example. In the “Understanding How Women Travel Report”, LA Metro revealed detailed findings on how women use their system. LA Metro released a specific “Understanding How Women Travel” survey in 2018 that reached 2,600 respondents \cite{lametro2019}. To supplement this data, LA Metro analyzed trip purposes using the National Household Travel Survey (an LA County subset) along with 6 other surveys not directly related to gender differences. More innovative forms of data collection contextualized the findings, including 3 focus groups, 100 hours of participant observations, and participatory workshops. These innovative forms of data collection contributed significantly to contextualizing experiences on transit for women, specifically lower-income women or non-English speaking women. The participatory workshops specifically were crucial in observing intersectionality, for example, how fully-employed women travel differently than part-time employed women. Much of the influence for the agency came from the City Woman’s Office in Vienna \cite{gender_mainstreaming} and \textit{Mobility of Care} concepts.

% What about a paragraph citing some of the studies that have used ODX data to tell us a lot about passenger travel (e.g. just not with a gendered lens?)
There is a growing body of work that uses ODX data to better understand rider travel behavior and make policy and operational improvements. Those use cases of ODX data range from understanding commuting patterns of transit users \cite{ma2017understanding}, and estimating trip purposes \cite{kusakabe2014behavioural} to linking customer delays to root causes \cite{doi:10.1177/03611981211054831}. Passenger segmentation, a task that involves clustering passengers into different user groups based on their common recurring travel behavior has also been a popular use of ODX data \cite{bhaskar2014passenger}. Unfortunately, fare card data rarely includes demographic information, which could be used to analyze travel patterns and fare burden by gender, among other attributes. This paper contributes to the literature by offering a novel large-scale study based in the United States on \textit{mobility of care} patterns using ODX data for a set of nearly 2 million users and tens of millions of trips that have taken place in the National Capital Region in the first quarter of 2019 (January 1\textsuperscript{st} - March 31\textsuperscript{st}).

% Perhaps add a paragraph about why survey data can be biased and unreliable? 

\section{Methodology}

\begin{figure}[!h]
\centering
\includegraphics[width= \textwidth]{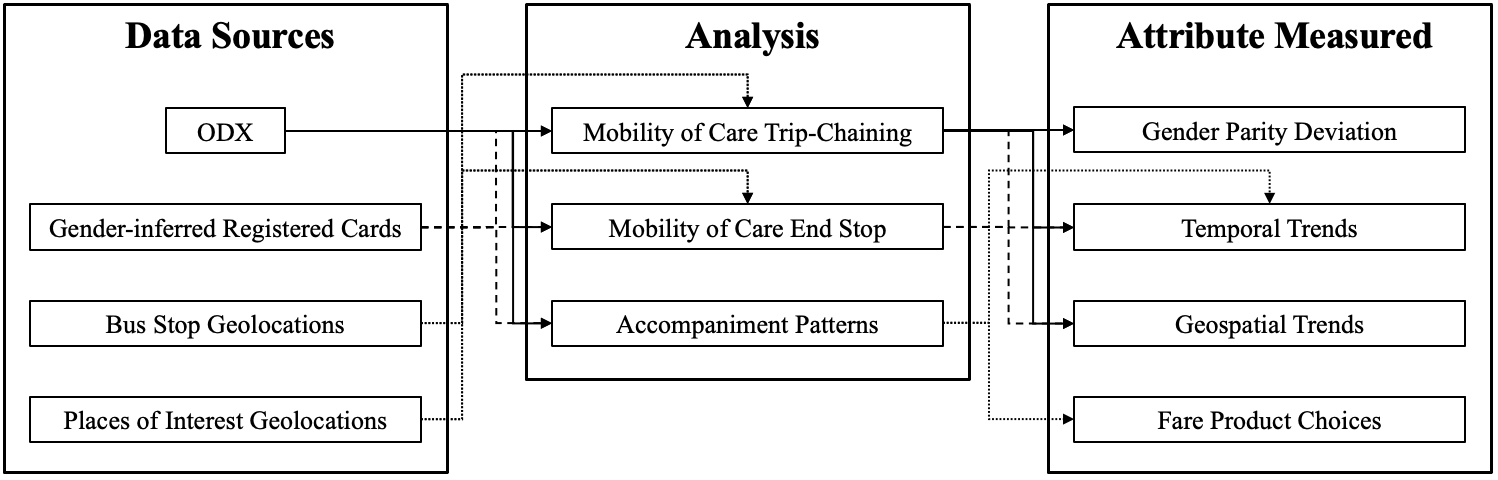}
\caption{\label{fig:analyses} Overview of data sources, analyses, and measured attributes.}
\end{figure}

\subsection{Origin-Destination and Registered SmarTrip Card Data}
The ODX dataset provided by WMATA contains the actual or inferred origin, destination, and transfer information for customers using the agency's SmarTrip card. This includes trip-level (journeys) and individual journey legs (stages) information such as boarding and alighting times and locations, routes used, distance traveled, and fare paid. The ODX data for the first quarter of 2019 contains 80,023,450 stages, 70,561,887 journeys, and 3,012,087 unique cards. Given that Metrorail stations are limited in number and serve larger areas compared to bus stops, many mobility of care and non-mobility of care destinations fall within the Metrorail station walksheds, making it harder to infer whether a given trip was to a \textit{mobility of care} destination. The walksheds for bus stops are much more compact. Therefore, we focus our geospatial analyses only on bus journeys to allow for a more granular analysis, where we can have higher confidence in origins and destinations in the proximity of bus stops. Of the aforementioned records, 25 million of all stages are bus stages, forming a total of 22 million journeys conducted by 814,726 users that have tapped into the Metrobus system during the first quarter of 2019.

Over 1.9 million customers in 2019 used a registered SmarTrip card (958K of whom were active in the first quarter), which asks for basic contact information, such as first and last name at registration. While the registered user customer data allows for conducting demographic and gender inference, it excludes cash transactions as it only includes tap-in/tap-out interactions with the WMATA system. Registration is also not mandatory, and registered users are more likely to be participating in the SmartBenefits (D.C's employer-based transit subsidy) program, receiving either a pre-tax, full, or partial transit subsidy from their employer. As a result, the registered card users dataset contains a higher share of customers who worked full-time and had typical pre-pandemic commute patterns. Hence, this study focuses on analyzing destinations where individuals conducting \textit{mobility of care} activities are expected to travel more, despite the higher likelihood of being on a 9-5 work schedule. Accompanying children to daycare centers and schools, and going to grocery stores are examples of places we expect this pattern of travel to be prevalent based on the literature, particularly during drop-off/pick-up times on weekdays.

\subsection{Gender Assignment to SmarTrip Cards}
% Copied from document that Awad sent 
WMATA has a dataset containing a limited number of customers ($\sim$5,000) who have provided detailed self-reported demographics, including gender. To conduct the large-scale analyses proposed for this project, it was necessary to acquire gender information for a larger number of customers. Gender inference based on the first name has grown in popularity in recent years, mainly due to the ease of acquiring large volumes of user names and self-reported genders through social media platforms. This has been studied by researchers including Liu \& Ruth \cite{liu2013s}, and Santamaría \& Mihaljevic \cite{santamaria2018comparison}. For baseline gender inference, we utilized the API integration for a commercial service, Genderize.io. Genderize scrapes open data from the internet, mainly public social media accounts, to collect counts and self-reported gender information associated with every first name. As of the time of writing, Genderize’s database contains gender counts and probabilities from over 114 Million entries (total entries in the database and not unique names). The API also provides the country breakdown associated with the names in their database.

\subsubsection{Pre-Processing and Gender Inference}
The data shared with the researchers only included first names and encrypted card IDs for the 1.9 million registered WMATA cards in $2019$ with no other personally identifiable information. Prior to running gender inference on the basis of the first name, the following steps were taken:
\begin{itemize}
    \item Names were pre-processed to remove blank entries and name fields reported as numbers.
    \item The remaining names were converted to Proper Form and white spaces were removed from the name string to remove discrepancies (hyphens are kept unchanged).
    \item A set of unique first names ($88,720$) were then taken and run through the Genderize API, retrieving the associated names, gender classification, the database counts for a given name, and the probability of gender classification.
    \item We saved those into a local database, which can be accessed to retrieve gender for registered users' mobility analyses. This step also reduces the effort for future name-to-gender inference to unique names as opposed to entirely new datasets.
    \item First names that remain with unidentified gender out of the Genderize API are then passed through the US Social Security Baby Names database, which lists the most popular 2,000 baby names in the US for every year from 1880 onwards \cite{nuessel2017note}.
\end{itemize}

Overall, a likely gender is associated with more than $60,000$ unique names, which were slightly more women-classified than men-classified. When joining the name-gender classification to the registered SmarTrip card data, we end with $961,118$ ($50.1\%$) men, $892,276$ women ($46.6\%$), and  $62,009$ cards with no associated gender classification ($3.3\%$). While the men-classified names had higher confidence associated with the gender prediction, the majority of both men's and women's classification probabilities are well above $90\%$. \ref{appendix:a} provides a validation using a WMATA survey dataset with self-reported demographic information. We compared the predicted to self-reported gender and found our method accurately inferred gender for 94\% of those customers. 

It is important to make explicit that this analysis does not seek to perpetuate the stereotype that those with names that are more likely to belong to women should be conducting \textit{mobility of care} work, or that a binary representation (female/woman, male/man) based on the name associated with sex-at-birth sufficiently represents people's gender identities, as demonstrated by D'Ignazio and Klein \cite{d2020data}. Instead, in the current state, studies in the literature consistently find that individuals identifying as women tend to perform a disproportionate share of care work. The purpose of this large-scale, name-based analysis is to aid in improving the system for caretakers and assessing the mobility patterns of those who are likely to be women-identifying individuals. This is only one way of doing so as the literature suggests and the available data permits.

\subsection{Geospatial Analysis}
\subsubsection{Sampling Registered SmarTrip Cards}
% post-processing of rider demographics
The dataset was balanced by inferred gender for the period of analysis (first quarter of 2019). The name-based gender inference yields more cards associated with men than women in the registered SmarTrip card dataset, although the survey data for WMATA consistently shows that more customers, particularly on the Metrobus system, are women. This discrepancy is attributed to the aforementioned bias of the registered SmarTrip users' dataset towards full-time employees, which in this case included slightly more men than women. To account for this discrepancy, we used all the women-classified cards and randomly drew a matching number of men-classified cards (hereinafter referred to as women and men). This results in a 50-50 baseline ratio of women to men in our analysis subset. Only the stages taken by the cards in this balanced random sample are included in the analysis. The analysis pipeline was re-run numerous times to ensure that this sampling process didn't introduce bias, and the results were found to be consistent. 

% Identifying POIs
\subsubsection{Identifying Places of Interest}

\begin{figure}[!h]
\centering
\includegraphics[width= 0.40\textwidth]{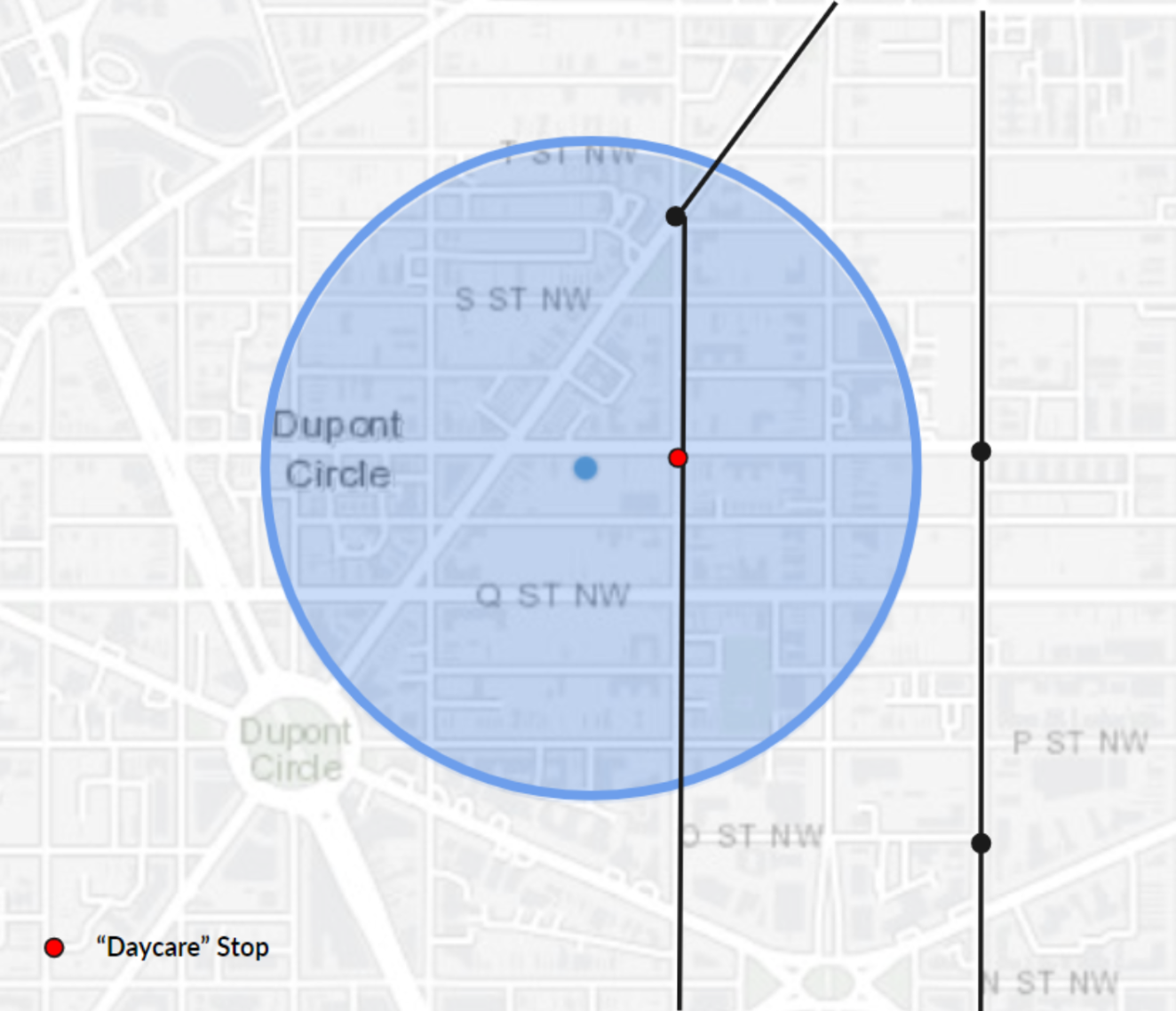}
\caption{\label{fig:poi_method} Identifying closest stop (red) to a POI (blue) per transit route-direction.}
\end{figure}

A geospatial approach was used to identify stages that potentially involve \textit{mobility of care}. If a rider boards or alights at a stop near a \textit{mobility of care} place of interest (POI), we can assume the journey (or part of it) was more likely to be a \textit{mobility of care} trip compared to stages at stops further from these POIs. We used the DC Open Data repository to identify the geographical location of daycares, schools, and grocery store locations in Washington DC and nearby areas in Maryland and Virginia served by the WMATA Metrobus system. One way of identifying stops near the POI is to simply include all stops within a certain walkable distance from the POI. However, depending on the distance criterion, this method may result in the inclusion of multiple stops along the same route and direction, while bus riders whose intention is to end up at that POI will most likely board or alight at the nearest stop along the route. We, therefore, implemented a route-direction-based algorithm to identify the stops serving a set of POIs.

The algorithm takes in a list of POI unique IDs and locations and the agency's GTFS feed for the time period of interest to identify a set of unique route-direction pairs. The route-direction pairs are generated by grouping all travel patterns that fall in the same route direction. Each unique route direction has a list of stops. For each POI, the algorithm iterates through those stops, identifies the nearest stop within a reasonable walking distance of up to 1/4 mile (400m) \cite{nabors2008pedestrian} to an individual POI (e.g. a daycare), and appends it to the set of stops closest to the particular POI class (e.g. all daycare centers in the area). The average city block distance in Washington DC is 528 feet (160m), so 400m provides a 1.5 block coverage before and after a certain POI. We found the average POI-to-stop distance identified by this algorithm to be 74 meters. This process is illustrated in Figures \ref{fig:poi_method} and \ref{fig:poi_stops}, and further elaborated on in \ref{appendix:b}.

\begin{algorithm}
\SetAlgoLined
  \caption{Identifying Bus Stops Closest to Places of Interest}
  \KwData{ODX data, WMATA bus stops, and places of interest geolocation coordinates.}
  \KwResult{Sets of unique nearest stops within 400m of a POI class for all bus routes.}
  %initialization;\\
  $POI \gets \{\text{daycares, schools, grocery stores}\}$\\
  $R \gets \{\text{all bus route-direction pairs}\}$\\
  $S \gets \{\text{all bus network stops}\}$\\
  \For {$p \in POI$}{
  $\text{Initiate POI class stops empty set } p_{stops}$ = \O \\
    \For {$r \in R$}{
        $\hat{s}$ = \O \\
        $d_{nearest} = \infty$ \\
        \For {$s \in S$}{
        \eIf{$d(p, s) \leq 400m \And d(p, s) < d_{nearest} \And s \notin p_{stops}$}
            {
            $\hat{s} = s$ \\
            $d_{nearest} = d(p, s)$ \\
            }
            {
            $continue$
            }
    }
    $p_{stops} \gets \hat{s}$ \\
    }
    }
    \Return $p_{stops}$ 
\label{algorithm:poi}
\end{algorithm}

\noindent Where $d(p, s)$ is the Euclidean distance between the geo-location of a POI and a bus stop.

\begin{figure}[!h]
\begin{subfigure}{.49\textwidth}
  \centering
  \includegraphics[width= 0.90\textwidth]{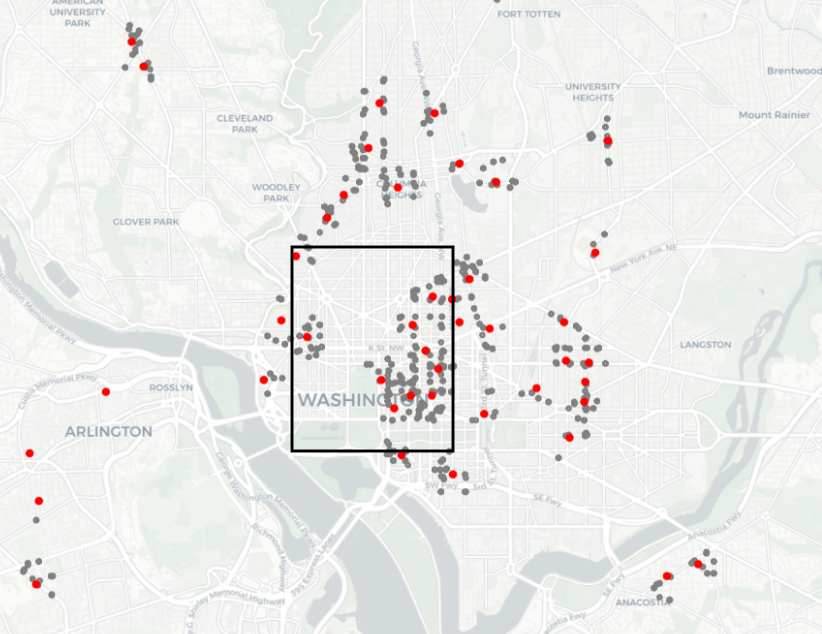}
  \caption{\label{fig:daycare_stops} Daycare stops.}
\end{subfigure}
\begin{subfigure}{.49\textwidth}
  \centering
  \includegraphics[width= 0.90\textwidth]{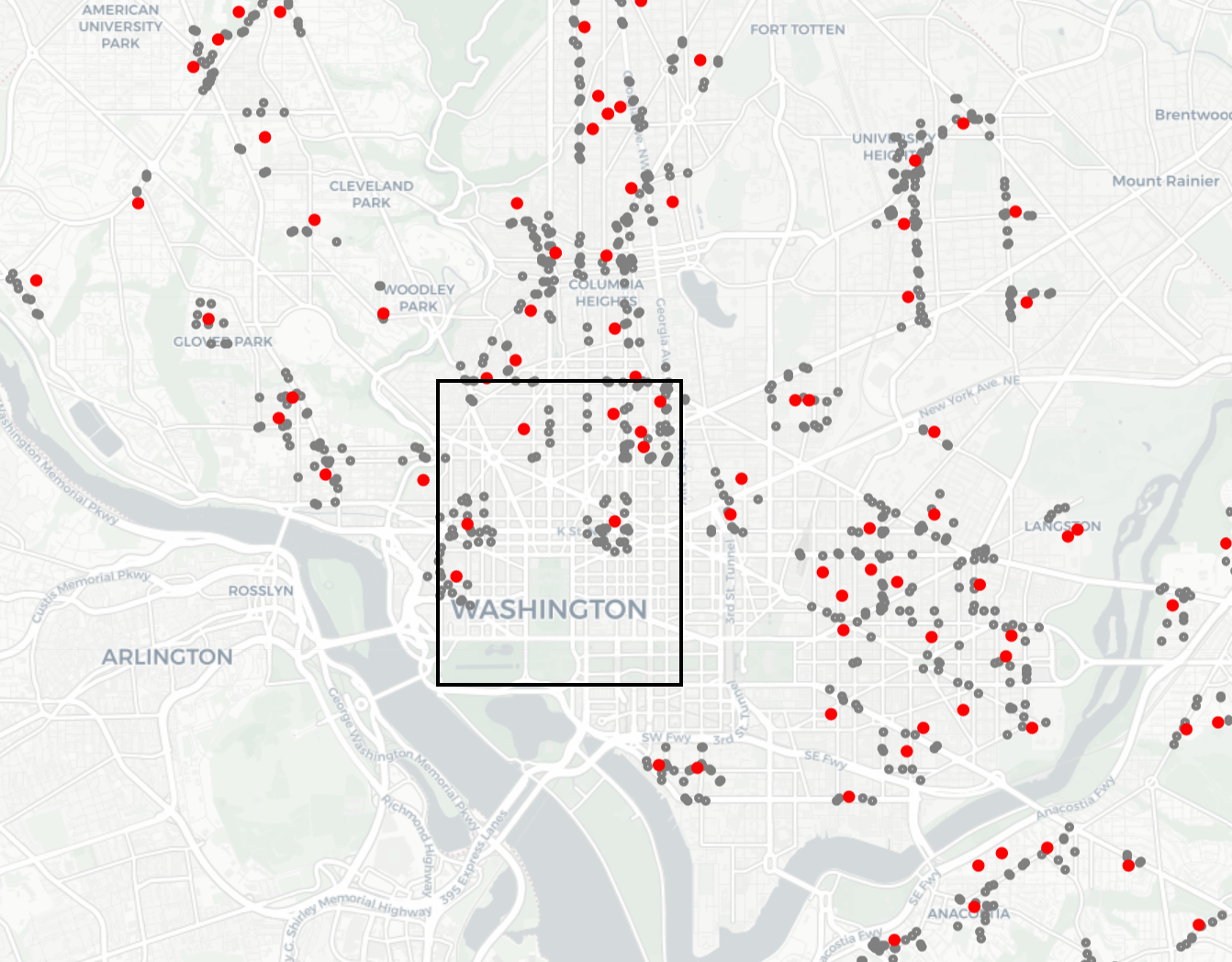}
  \caption{\label{fig:school_stops} School Stops.}
\end{subfigure}
\caption{\label{fig:poi_stops} Identified nearest stops (in grey) for DC Schools and Daycares (in red). The city center boundary is depicted with a black box.}
\end{figure}

\subsubsection{Identifying Mobility of Care Travel Pattern}
To help identify travel patterns more likely to relate to \textit{mobility of care} while reducing potential noise, we make the assumption that the anticipated journey characteristics would include regular (recurring) travel patterns, more evident on weekdays and peak in the morning and afternoon. We define regularity in system usage by limiting the analysis to SmarTrip cards that were active on the system for 10 days or more during the period of the study. After limiting the data to active users, 55 million journeys by 782,009 unique SmarTrip cards remain. This suggests that approximately 80\% of the journeys are taken by the top 26\% of active users. Of those active-user records, 17 million stages and 15 million journeys took place on Metrobus. Table \ref{tab:data_processing} shows the data reduction along the processing steps. The analysis presented hereafter is for 8.55 million stages and 7.67 million journeys conducted in the first quarter of 2019 by 215,000 active, registered SmartTrip cards where gender could be inferred (equally sampled to represent 107,500 women and men). 

% \begin{enumerate}
%     \item Regular, recurring travel patterns,
%     \item More evident on weekdays, and
%     \item Peak in the morning and afternoon.
% \end{enumerate}

\begin{table}[!h]
 \renewcommand{\arraystretch}{1.0}
  \centering
  \caption{Dataset Processing Results}
    \begin{tabular}{lccc}
    \toprule
    & {\textbf{Unique Cards}} & {\textbf{Total Journeys}} & {\textbf{Total Stages}} \\
    & (thousand) & (million) & (million) \\
    \midrule
    \textbf{Full 2019 Q1 Dataset} & 3,012 & 70.56 & 80.02 \\
    \midrule
    \textbf{10+ Active Days} & 782   & 55.40 & 64.12 \\
    \midrule
    \textbf{On Metrobus} & 420   & 15.65 & 17.78 \\
    \midrule
    \textbf{Registered w/ Gender Inference} & 215   & 7.67  & 8.55 \\
    \bottomrule
    \end{tabular}%
  \label{tab:data_processing}%
\end{table}%

In order to identify travel behavior that is more likely to be associated with a specific POI, the first case we analyzed includes only journeys that have multiple stages and where the second boarding or higher occurs at a \textit{mobility of care} POI stop. In other words, it only includes journeys where someone trip-chained near a POI. An example of a journey included in this setup is when someone takes the bus to drop a child off at daycare and, shortly after, boards another bus to continue to their final destination.  The identification of \textit{mobility of care} trips was initially restricted to the multi-stage journeys to reduce the likelihood that we capturing scenarios where the rider consistently enters the system for the first time at a stop near a POI, for a reason unrelated to the POI (e.g. the POI happens to be near their home or work).

The second case we analyzed relaxes the initial trip chaining limitation, including any journey that has a stage that ends near a POI.  However because Metrobus customers tap in to board, but do not tap out, the ODX process is only able to infer alighting location for 65\% of bus stages. This second case allows us to test if we can identify \textit{mobility of care} trips by limiting the data to stages with an alighting stop inference, but without limiting it to the subset of trips where we observe a trip chaining behavior.

\noindent To summarize the travel patterns we analyze:

% 	\begin{itemize}
% 	    \item ...$ \rightarrow Bus \rightarrow POI \rightarrow Bus \rightarrow$...: observed by standard analysis
% 	    \item ...$\rightarrow Bus \rightarrow POI \rightarrow$...: observed by end stop analysis
% 	    \item $POI \rightarrow Bus \rightarrow$...: not observed
% 	\end{itemize}
	\begin{itemize}
	    \item Case 1 (baseline analysis): Origin $ \xrightarrow[]{\text{Bus}}  POI \xrightarrow[]{\text{Bus}} $ Destination
	    \item Case 2 (end stop analysis): Origin $\xrightarrow[]{\text{Bus}} $ POI
	    %\item $POI \rightarrow$ Destination: not observed.
	\end{itemize}

Ultimately, we observe the deviation of the ratio of women riders from gender parity over the course of the day at the stops identified closest to the \textit{mobility of care} destinations. Starting from the assumption that gender should play no role in mobility behavior, the ratio of customers at any stop should not significantly deviate from gender parity induced by the equal sampling of the gender-labeled dataset of users. The gender ratio was calculated by taking the total boardings (alightings in the case of the end stop analysis) at POI stops and all bus network stops, grouped by 15-minute intervals from 6 AM to 10 PM. Stages that started within this time period were grouped by unique SmarTrip card ID and the percentage of women-classified cards was calculated. For simplicity going forward, we refer to results generated by the women-classified cards as "women" and by the men-classified cards as "men".  

We define the deviation from gender parity as follows. For each 15-minute interval from 6 AM to 10 PM, $t \in \{$(6:00 - 6:15), (6:15 - 6:30), (6:30 - 6:45), ..., (21:45 - 22:00)$\}$, we calculate the mean ($\bar{x}_t$) and standard deviation ($\bar{\sigma}_t$) of the deviation from gender parity across all origin-destination pairs. $n_t$ is the number of trips in time interval $t$. 

$$\textit{Deviation From Gender Parity}_t = \bar{x}_t = \frac{\textit{\# Trips made by women in time t}}{\textit{Total number of trips in time t}} - 50\%$$

$$\textit{95th Percentile CI}_t = \bar{x}_t \pm 1.96 \left( \frac{\sigma_t}{\sqrt{n_t}} \right)
$$

\subsection{Accompaniment Analysis}

In addition to the geospatial approach for inferring \textit{mobility of care}, we conducted a mode-and-destination-agnostic analysis to identify SmarTrip cards that exhibit "accompaniment patterns". Many forms of mobility of care tripmaking involve one person accompanying a child to school or an elderly relative to a doctor's appointment, for example. This accompaniment behavior manifests in SmarTrip taps of two cards in rapid succession on a regular basis, with one of the cards being a student, senior, or disabled fare product. Certain types of mobility of care trips such as grocery shopping or daycare trips, where the child is too young to have their own SmarTrip card, might not be captured through such an accompaniment analysis. However, such analysis does allow us to infer the \textit{mobility of care} trips on Metrobus as well as Metrorail, and it is not tied to a specific geographic point of interest. To identify accompanying cards we: 

\begin{enumerate}
    \item Include all 55 million journeys (and their 64 million stages) conducted by active WMATA users in the first quarter of 2019.
    \item Specify fare products of interest. As opposed to exhaustively identifying all cards that may travel together, we optimize an algorithm to search for cards that exhibit a recurring pattern of accompanying WMATA's student, senior, and disability fare products.
    \item We define an accompaniment pattern as one where a SmarTrip card with one of the aforementioned fare products is observed tapping into the WMATA system (at buses or Metrorail stations' fare gates) recurrently accompanied by another card (of any fare product) at least 4 times in a month (i.e. two round trips) or 10 times through Q1 of 2019.
    \item Tap-ins have to be consecutive (regardless of whether the accompanying or accompanied taps-in first), take place 30 seconds apart at most, and are at the same Automatic Fare Collection (AFC) machine for Metrobus tap-ins, and mezzanine-level (station gate) for Metrorail.
    \item Accompaniments are classified as student, senior, or disabled based on the fare product held by the accompanied card.
\end{enumerate}

Although there are no specific fare products for infants and toddlers (who would be accompanied to daycares) or associated with grocery shopping, this supplementary analysis covers other facets of \textit{mobility of care} that we haven't explored through the geospatial approach. 

\section{Results and Discussion}
\begin{figure}[!h]
\centering
\includegraphics[width= .9\textwidth]{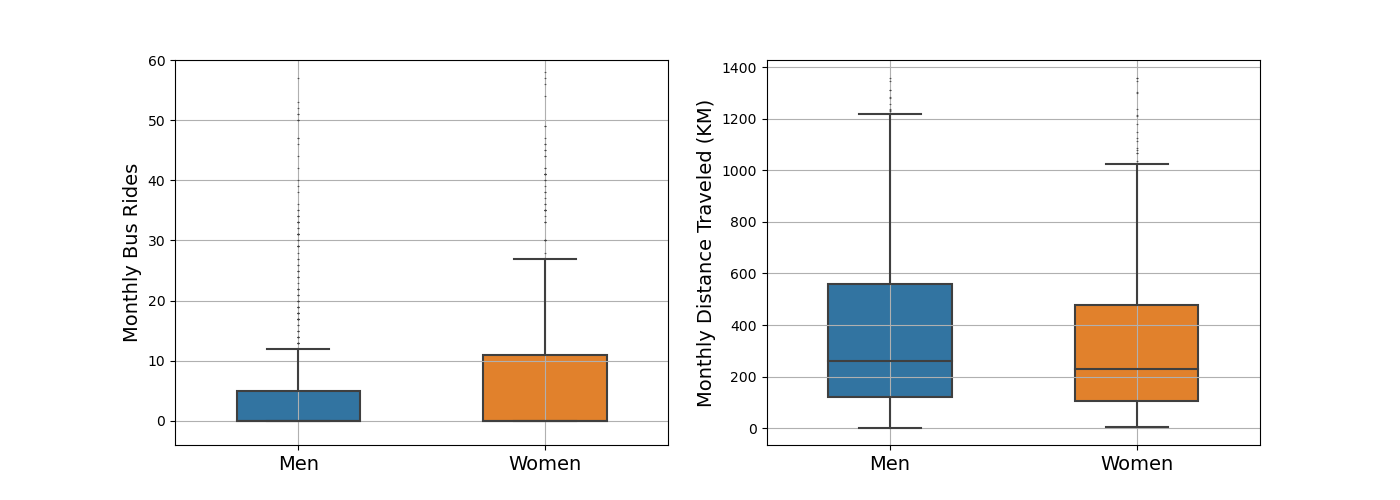}
\caption{\label{fig:distributions} Distributions of observed bus journeys and journey lengths by gender. The box shows the first quartile, median, and third quartile. The “whiskers” extend to points that lie within the 1.5 inter-quartile range of the lower and upper quartile. }
\end{figure}

Looking at the big-picture differences between how men and women use the system, we found the results consistent with previous studies on the topic. Using only the inferred gender as a differentiator for all the registered SmartTrip cards in the first quarter of 2019, we found that women are more likely to take the bus and travel shorter distances, as illustrated in Figure \ref{fig:distributions}. Daily and weekly patterns were consistent with the monthly distributions. While those findings are in line with similar studies on the topic, we did not find conclusive evidence of other predominant behaviors found in the literature, which include women traveling more often during off-peak hours, and conducting journeys that involve more transfers. This was not unexpected, given the aforementioned bias that the registered SmarTrip card data is more likely to over-represent full-time employees with a typical pre-pandemic commuter pattern.

\subsection{Places of Interest Analysis}
We chose to observe the gender distribution of a subset equally sampled by gender of active transit users to the \textit{mobility of care} destinations, with the knowledge that per the literature, journey characteristics associated with \textit{mobility of care} were found consistent among women with full-time jobs and otherwise. We expected women to travel to and from stops near these POIs at larger rates than men and found a significant difference in the percentage of women boarding a second stage or higher (after a presumed visit to a \textit{mobility of care} POI) at stops near POIs compared to men. This is quantified by looking at the deviation from gender parity, given that with an equally balanced sample (based on inferred gender) the neutral trip distribution should be 50\% for each gender. 

Although the sample of cards was balanced, women took 2.1\% fewer trips than men on average. However, we observe that women made 1.5\% to 1.8\% more trips for daycare, school, and groceries compared to men at the median bus stops serving \textit{mobility of care} places of interest. This gap widens significantly at the higher end of the distribution, with women being at least 10\% more likely to conduct \textit{mobility of care} trips than men at the 90th percentile. These results are shown in Table \ref{tab:overall_stats}.  This indicates that women are significantly more likely to be trip chaining near POIs associated with \textit{mobility of care} travel. While not all transit users observed transferring near a \textit{mobility of care} POI would eventually be patronizing it, the higher ratio of women observed in those particular places of interest does indicate a higher probability of women conducting \textit{mobility of care} work.

\begin{table}[!h]
\centering
\renewcommand{\arraystretch}{1.0}
\caption{Statistics for Women Ratio Deviation From Gender Parity}\label{tab:overall_stats}
\begin{tabular}{lcccc}\toprule
&{\textbf{25\textsuperscript{th} Percentile} \par} &{\textbf{50\textsuperscript{th} Percentile} \par} &{\textbf{75\textsuperscript{th} Percentile} \par} &{ \textbf{90\textsuperscript{th} Percentile} \par}\\\cmidrule{2-5}
\textbf{Daycares} &-5.70\% &1.80\% &9.10\% &15.00\%\\
\textbf{Schools} &-4.50\% &1.50\% &6.70\% &11.30\% \\
\textbf{Grocery} &-5.90\% &1.80\%	&7.70\%	&12.50\%  \\
\textbf{All} &-4.58\% &-2.10\% &0.00\% &1.50\%  \\
\bottomrule
\end{tabular}
\end{table}

\begin{figure}[!h]
\centering
\includegraphics[width= .6\textwidth]{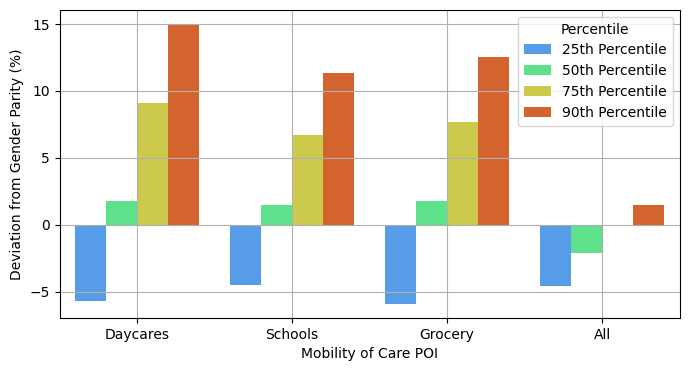}
\caption{\label{fig:poi_dist} Deviation from gender parity for trips conducted by women across Mobilit of Care POIs. }
\end{figure}

\newpage
Table \ref{tab:all_stats} shows counts of POI stops and the total and the average number of women and men trip-chaining stages observed. The closest bus stop to the POI is a proxy of travel to that location, but we do not know from the card-based data that someone actually visited a \textit{mobility of care} POI. We performed an additional check using only the data outside the city center to test if the results were being driven by the density of the development downtown. Evaluating stops that fall outside the city center, we find consistent results as demonstrated in Table \ref{tab:outside_stats}. The total number of stages includes all stages for the first quarter of 2019. We find that on average, during every 15-minute interval, there is one additional woman that boards at stops near daycare centers compared to men, despite having slightly fewer women stages on the system's average. We observe this discrepancy across all \textit{mobility of care} POIs, including an average of 3 additional women boardings per hour at stops near grocery stores.

\begin{table}[!h]
\centering
\renewcommand{\arraystretch}{1.0}
\caption{Statistics For Weekday Stages Near POIs At All Stops}\label{tab:all_stats}
\begin{tabular}{lccccccc}\toprule
&{$\#$ \textbf{Stops} \par} &{$\Sigma$ \textbf{Women Stages} \par} &{$\Sigma$ \textbf{Men Stages} \par} &{$\overline{\textbf{Women/hr}}$ \par} &{$\overline{\textbf{Men/hr}}$ \par} &{\textbf{$\Delta$} \par} \\\cmidrule{2-7}
\textbf{Daycares} &594 &95,857 &86,270 &10.1 &9.1 &1 \\
\textbf{Schools} &1,029 &140,599 &126,692 &8.5 &7.7 &0.82 \\
\textbf{Grocery} &288 &92,210 &80,823 &20.0 &17.5 &2.5 \\
%\textbf{Total} &12,585 &793,856 & &0.99 & & \\
\bottomrule
\end{tabular}
\end{table}

\begin{table}[!h]
\centering
\renewcommand{\arraystretch}{1.0}
\caption{Statistics For Weekday Stages Near POIs Outside City Center}\label{tab:outside_stats}
\begin{tabular}{lccccccc}\toprule
&{$\#$ \textbf{Stops} \par} &{$\Sigma$ \textbf{Women Stages} \par} &{$\Sigma$ \textbf{Men Stages} \par} &{$\overline{\textbf{Women/hr}}$ \par} &{$\overline{\textbf{Men/hr}}$ \par} &{\textbf{$\Delta$} \par} \\\cmidrule{2-7}
\textbf{Daycares} &394 &61,103 &54,972 &9.7 &8.7 &1 \\
\textbf{Schools} &870 &119,951 &109,328 &8.6 &7.8 &0.8 \\
\textbf{Grocery} &238 &81,593 &69,414 &21.4 &18.2 &3.2 \\
\bottomrule
\end{tabular}
\end{table}

\subsubsection{Daycares}
\begin{figure}[!h]
\begin{subfigure}{.49\textwidth}
  \centering
  \includegraphics[width= \textwidth]{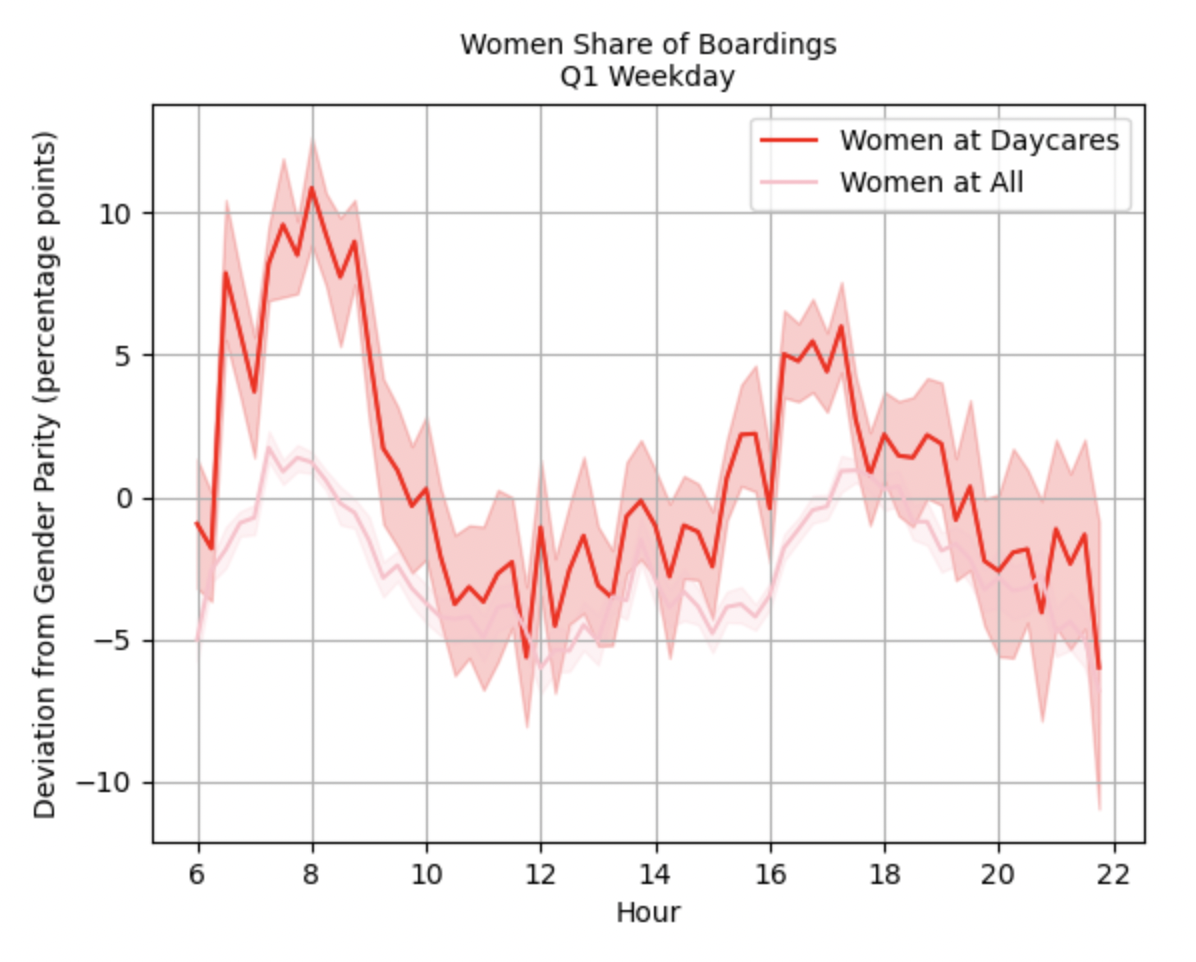}
  \caption{\label{fig:daycare_ratio} All daycare stops.}
\end{subfigure}
\begin{subfigure}{.49\textwidth}
  \centering
  \includegraphics[width= \textwidth]{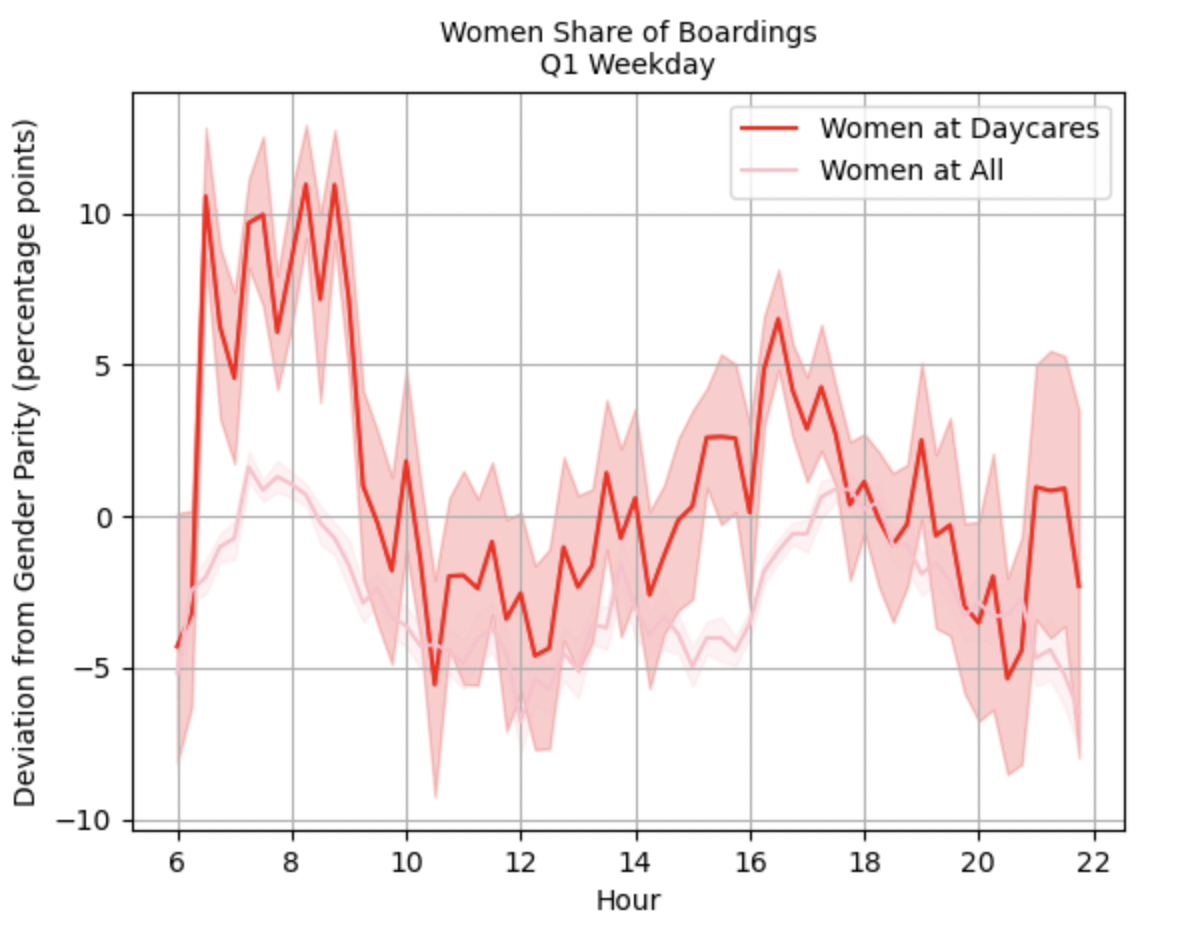}
  \caption{\label{fig:daycare_ratio_outside} Daycare stops outside city center.}
\end{subfigure}
\caption{\label{fig:daycare_ratios} Gender ratio deviation from 50\% women during weekdays near daycares.}
\end{figure}

Next, we observe the changes in gender ratio distribution across the day. Figure \ref{fig:daycare_ratios} illustrates the average deviation from gender parity near daycare stops (6\% of all stops) compared to women's trip share at all WMATA Metrobus stops, which serves as a baseline value for comparison. The upper and lower envelopes illustrate the 95\% confidence interval of the women ratio deviation from gender parity across different stops and days in the first quarter of 2019. 

We observe an average of 10\% more women (60\% of riders are women) trip-chaining at stops near daycares during AM drop-off times compared to men. At the same time of day for all destinations, women take 52.5\% of trips, only 2.5\% above parity, a 7.5\% difference with trips starting or ending near daycares. Since daycare stops make up only approximately 4\% of all stops, there is limited interaction between these two lines. The higher deviation during peak commute hours is expected as this data is skewed toward full-time employees, it is expected that full-time employed women are more likely to board at a daycare stop following a drop-off or pick-up respectively during the AM and PM peak hours. We observe a minimal difference in the gender ratio during the middle of the day between POI stops and all stops. This supports the hypothesis that women are more likely to exhibit the daycare/school drop-off/pick-up travel pattern during particular hours of the day and that there is not a consistent geographical correlation with the stop locations across all times. We also observe more spread in the pick-up times in the afternoon while the AM drop-off window is narrower. This suggests women traveling to pick-up may choose to do so over a wider spread of time, or that women and men may do a more equal share of pick-ups. To account for the geographical interaction between this observed deviation and city center dynamics, we evaluate \textit{mobility of care} stops outside the city center. We defined the Washington DC city center as the area bounded by Independence Ave SW \& U Street NW, and 6th Street NW \ and 23rd Street NW. The trends observed were found to be consistent for daycare stops outside the city center.  We also observe more variation in the women ratio during the afternoon and evening for stops outside of the city center, illustrated by the wider confidence intervals. This is attributed to the lower number of people boarding from those stops later in the day, leading to small changes in ridership to cause higher volatility.

\begin{figure}[!h]
\centering
\includegraphics[width= .55\textwidth]{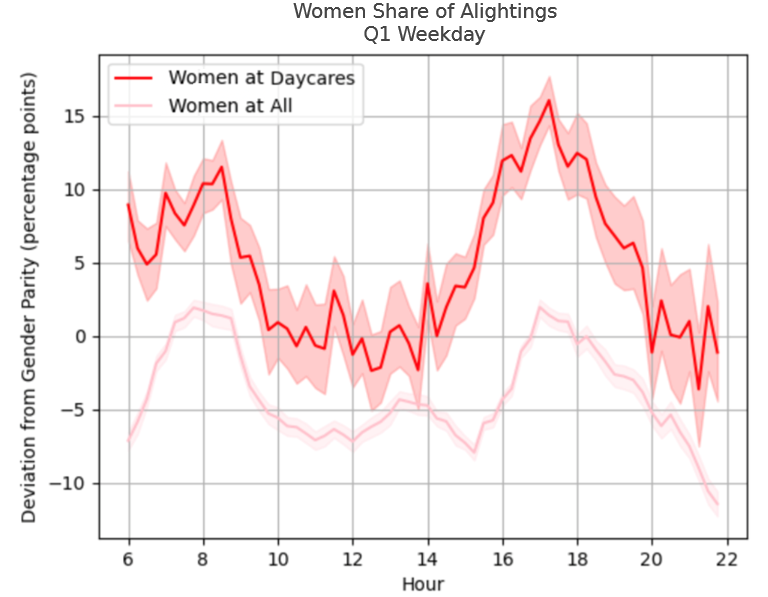}
\caption{\label{fig:daycares_ratio_end}  Gender ratio deviation for women alightings during weekdays at daycare stops.}
\end{figure}

Looking at the second case using end stop analysis that includes the  65\% of the data where the alighting stop of a stage was inferred, we observed a similar trend, with the women ratio peaking at commute times. We find a consistent trend in the AM peak, while in the PM peak, we observe 15\% more women ending their stages or journey at a stop near a daycare center. While a possible explanation is that more people live near daycares, given that the baseline is a gender-balanced sample it is unlikely this deviation can be entirely explained by geographical association. We don't expect women to disproportionately arrive at home stops more often than men. Thus, this may be able to capture women who walk or utilize a different mode of transport to drop off or pick up their children at daycares before taking transit for their subsequent stages.

\subsubsection{Schools}

\begin{figure}[!h]
\begin{subfigure}{.49\textwidth}
  \centering
  \includegraphics[width= \textwidth]{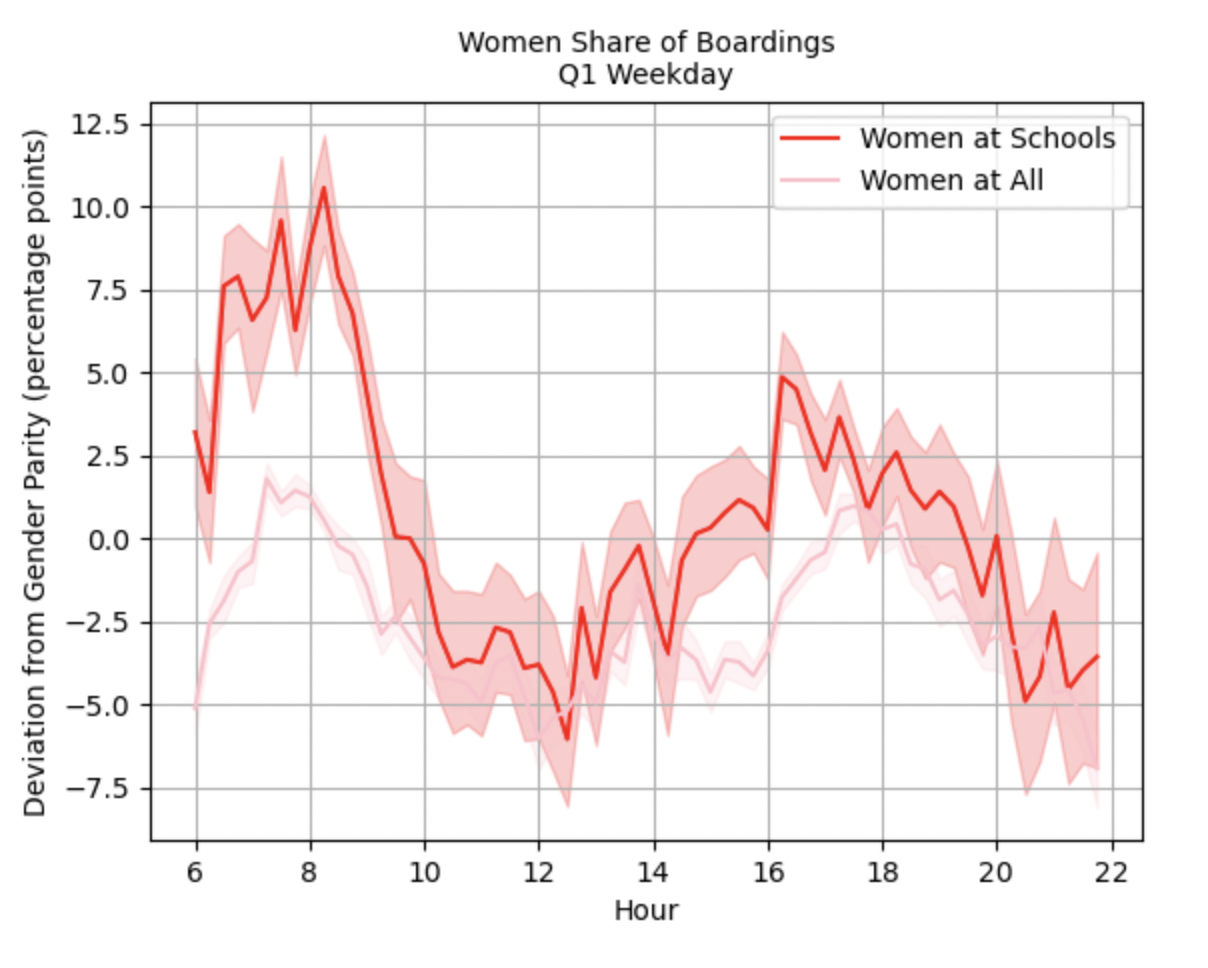}
  \caption{\label{fig:schools_ratio} All school stops.}
\end{subfigure}
\begin{subfigure}{.49\textwidth}
  \centering
  \includegraphics[width= \textwidth]{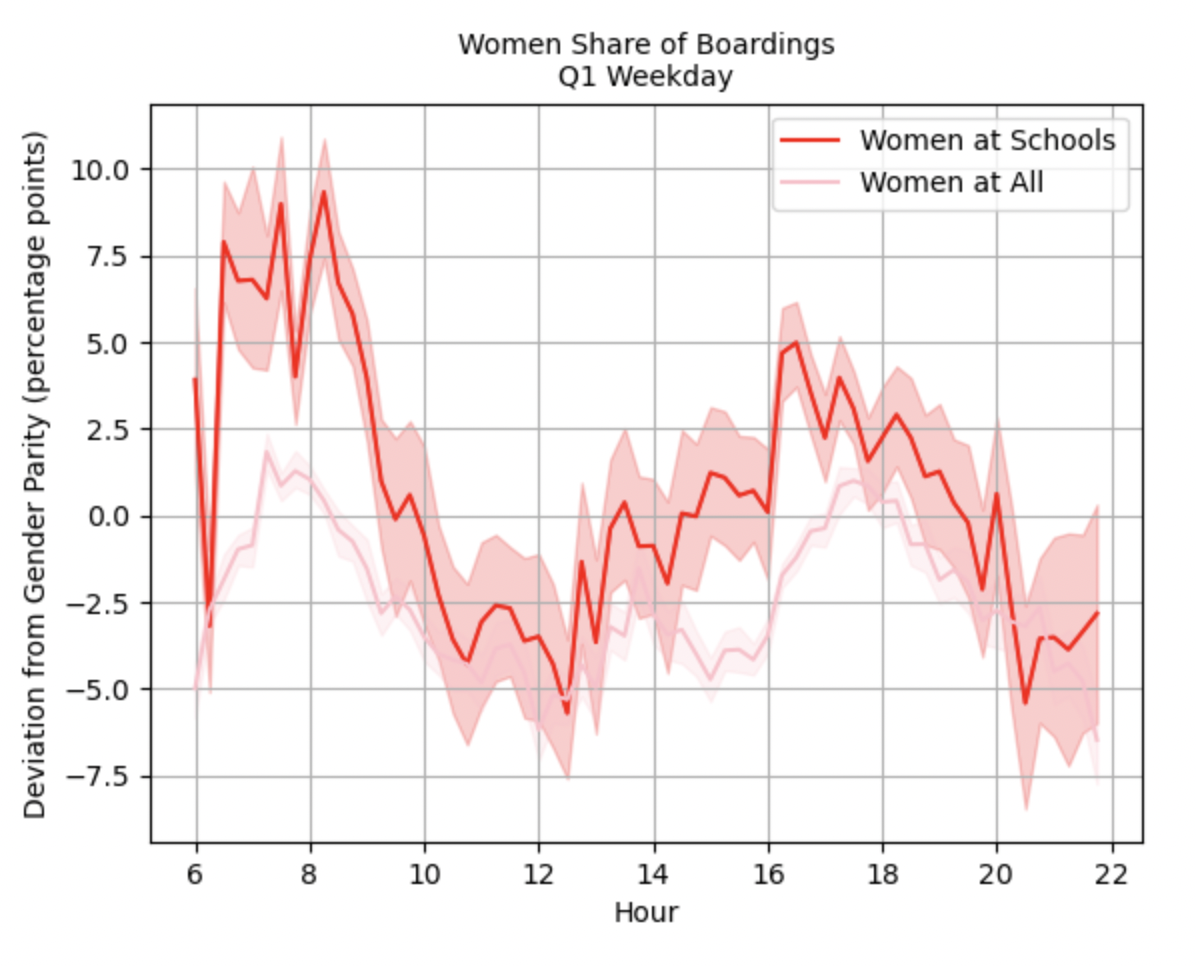}
  \caption{\label{fig:schools_ratio_outside} School stops outside city center.}
\end{subfigure}
\caption{\label{fig:schools} Gender ratio deviation from 50\% women during weekdays near schools.}
\end{figure}

We observe a similar trend at stops near schools (8\% of all stops): there are approximately 10\% more women than men who board a subsequent stage after a presumed drop-off during peak hours. Similar to what we observed at daycares, the trend holds for school stops outside the city center, as well as when observing journey stages that end near school stops. One noticeable difference is that there is a narrower AM drop-off window with a steep decline beginning at 8 AM, which is the time when school starts for DC Public School students. Hence, this trend is extremely logical, which further validates that our workflow for identifying frequent riders and nearest stops to places of interest is successful. 

\subsubsection{Grocery Stores}
\begin{figure}[!h]
\begin{subfigure}{.49\textwidth}
  \centering
  \includegraphics[width= \textwidth]{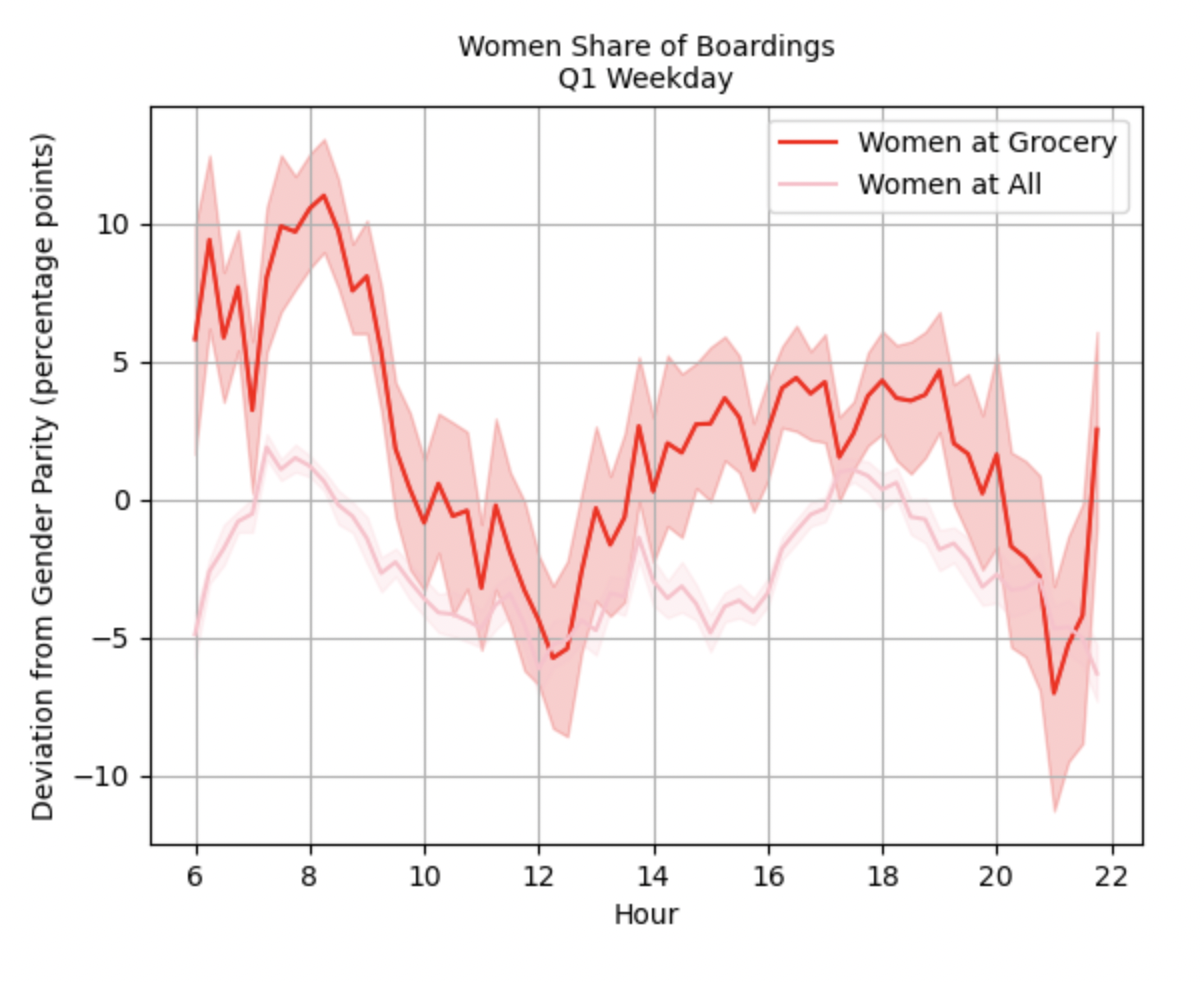}
  \caption{\label{fig:grocery_all} All grocery store stops.}
\end{subfigure}
\begin{subfigure}{.49\textwidth}
  \centering
  \includegraphics[width= \textwidth]{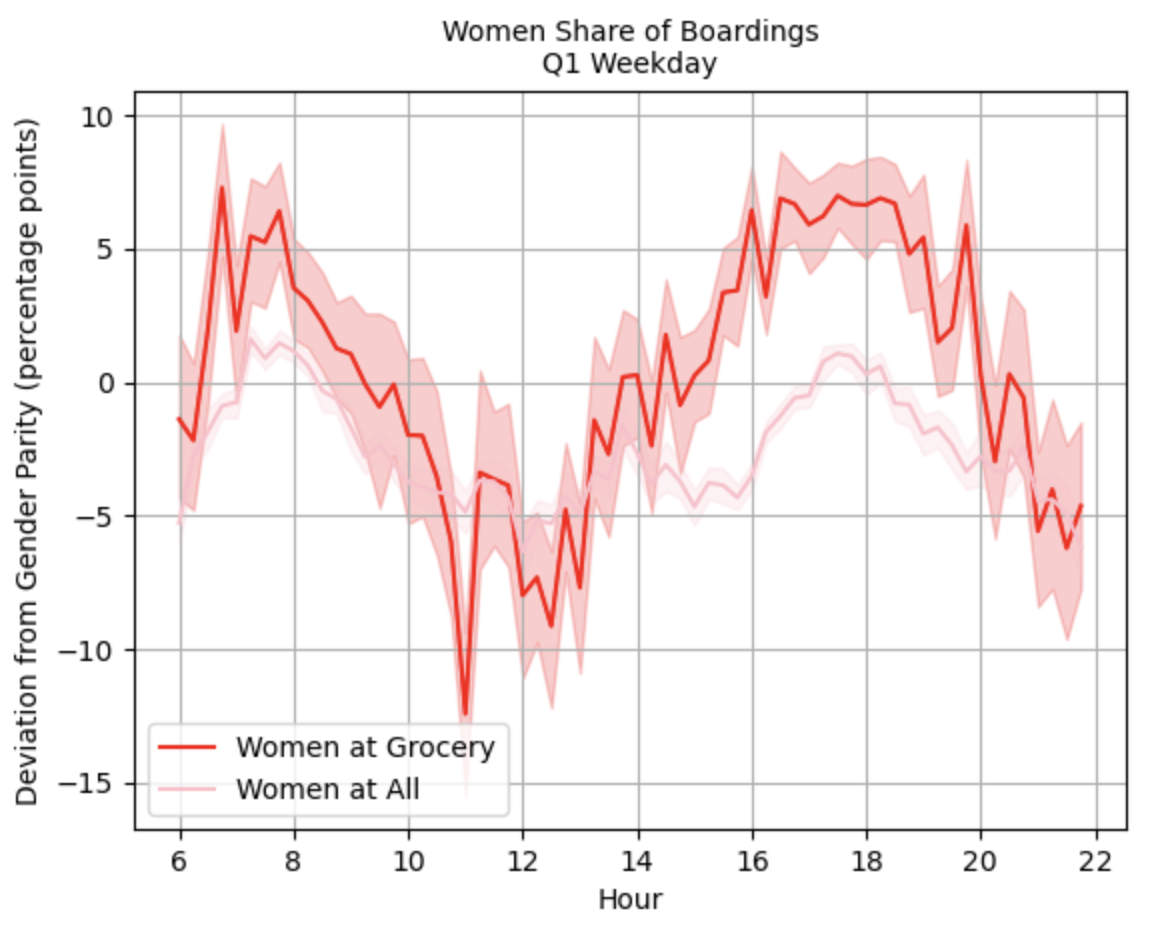}
  \caption{\label{fig:grocery_outside} Grocery stops outside city center.}
\end{subfigure}
\caption{\label{fig:grocery} Gender ratio deviation from 50\% women during weekdays near grocery stores.}
\end{figure}

We observe a similar trend at grocery stores  (2\% of all stops). The confidence intervals, however, are generally wider, particularly in the afternoon suggesting higher variability. This may be attributed to grocery store locations typically being in areas with much higher mixed use as compared to daycares and schools. Evidence of this is that when we look at grocery store stops outside the city center illustrated in Figure \ref{fig:grocery_outside}, we see that there is less of a relationship between grocery store stops and an increase in the ratio of women travelers in the morning, where the trend is closer to the overall trend. However, the observed higher percentage of women during the PM hours is robust, as well as when we observe trips that end at stops near grocery stores illustrated in Figure \ref{fig:grocery_end}, all with significantly tighter confidence intervals. It may also be because visits to grocery stores do not have set hours the way that daycare and school pick-up and drop-offs do.

\subsection{Comparing Alternative Travel Patterns}
We found the results for end-stop (alighting) to be consistent with the trip-chaining scenario in schools and grocery stores, similar to what was previously illustrated for daycares. For brevity, we provide these graphs in \ref{appendix:c}. The gender ratio deviation plots for school and grocery alightings are illustrated in Figure \ref{fig:school_grocery_endstops}. Additionally, we assessed and compared the change in the ratio of women near \textit{mobility of care} stops during the weekends. Overall, we observe significantly fewer women on the system on weekends, an average decrease of -10\%. This is reflected by the overlapping trend lines for all stops and POI stops illustrated in Figure \ref{fig:weekend}. This suggests that women conducting \textit{mobility of care} trips are not more likely to trip-chain in the vicinity of those POIs outside weekdays at the hours we hypothesized, which further validates our results. A logical exception to this, however, is observed at the POI stops near grocery stores, where more women can be observed on weekends, despite the significant decrease in women riders overall on weekends. This is illustrated in Figure \ref{fig:grocery_weekend} of the appendix with the average grocery store stops at gender parity and ones falling in the higher percentiles having an increase of over 10\% women riders during the afternoon and early evening periods.

\subsection{Analysis of Accompaniment Patterns}

\begin{figure}[!h]
\begin{subfigure}{.49\textwidth}
  \centering
  \includegraphics[width= \textwidth]{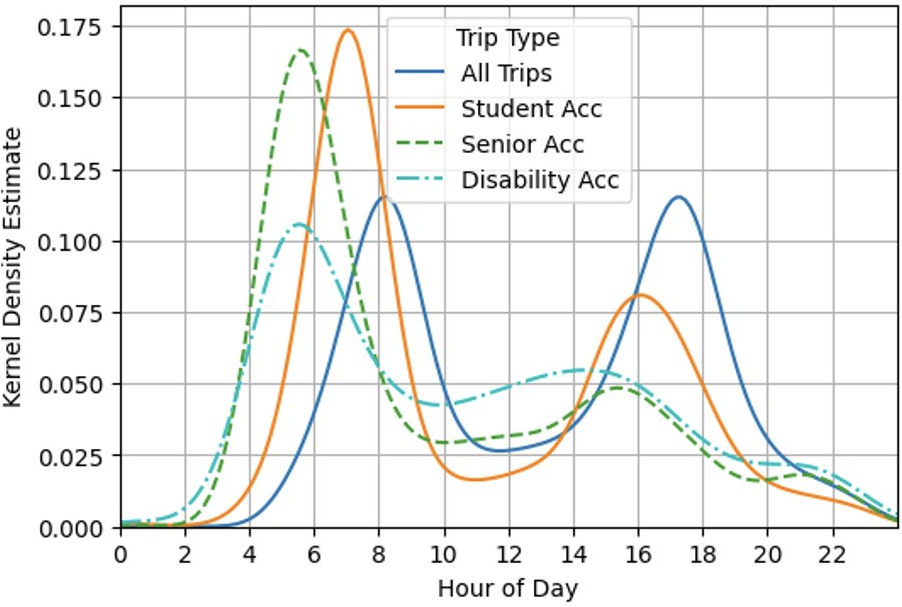}
  \caption{\label{fig:acc_weekday} Weekdays.}
\end{subfigure}
\begin{subfigure}{.49\textwidth}
  \centering
  \includegraphics[width= \textwidth]{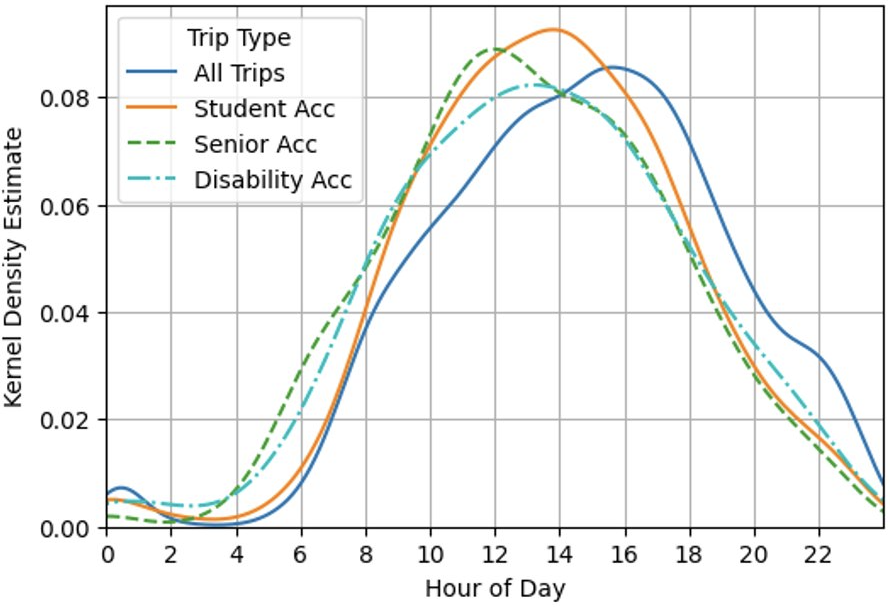}
  \caption{\label{fig:acc_weekend} Weekends.}
\end{subfigure}
\caption{\label{fig:acc_distribution} Probability distribution of different accompaniment types by hour of day.}
\end{figure}

In the first quarter of 2019, we identified 500K journeys to be \textit{mobility of care} by accompaniment. Different types of accompaniments were clearly differentiable based on the time of day in which they were more likely to occur. We found that student accompaniments are significantly more likely to occur around pick-up and drop-off times, and the least likely trip type to take place during the day. Senior and disabled individuals' accompaniment was found to start earlier in the day compared to other trips (student accompaniment or general/commuting trips). Additionally, they are also more likely to take place during the off-peak compared to other types of trips. The probability density functions of those trips by time of day are illustrated in Figure \ref{fig:acc_distribution}.

Our geospatial analysis of \textit{mobility of care} identified $\sim$4\% of all trips (2 million out of 54 million multi-modal journeys made by active users) as \textit{mobility of care} trips. For SmarTrip cards that we identified as ones exhibiting an accompaniment pattern, nearly 25\% of the trips conducted are also identified as \textit{mobility of care} trips through the POI-based analysis. In contrast to the very reasonable temporal patterns associated with those accompaniment patterns during the weekday hours, we find that there are no consistent patterns for those trips during the weekends, which further validates those results. This can be observed in Figure \ref{fig:acc_weekend}, where all types of weekend trips follow an almost identical distribution of taking place late AM through mid-day.

\begin{figure}[!h]
\centering
\includegraphics[width= .50\textwidth]{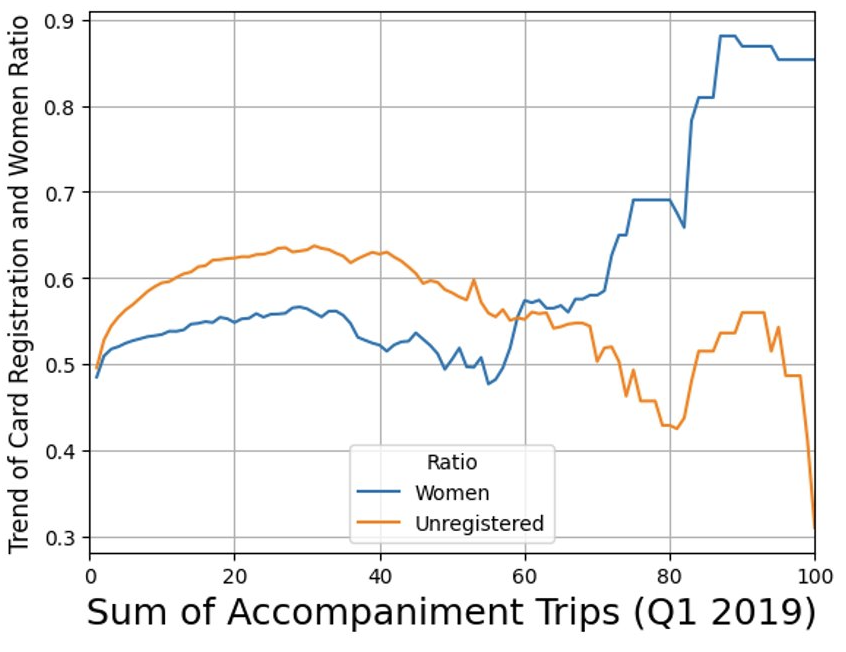}
\caption{\label{fig:acc_ratios} Ratios of unregistered cards and registered cards classified as women.}
\end{figure}

\newpage
Finally, for accompanying cards that were registered customers and had a name-based gender inferred, found that the likelihood of an accompanying card belonging to a woman significantly increases as the total number of accompanying trips increases. We also find that the percentage of unregistered cards significantly decreases at higher accompaniment rates. Both trends are illustrated concurrently in Figure \ref{fig:acc_ratios}. The latter trend is attributable to customers conducting high accompaniments switching from the general pay-as-you-go full-fare product to fare products that accommodate high journey and trip chaining rates (e.g. weekly passes and specific products for those accompanying the elderly and/or disabled). More details can be seen in \ref{appendix:d}.

\subsection{Policy Implications}
Seemingly gender-neutral or gender-agnostic policies can have significant impacts on women. This study adds to the growing body of research demonstrating that women and men do not follow the same tripmaking behaviors, and in particular, that women are more likely to use public transit to conduct \textit{mobility of care} trips. Regardless of the gender of the person performing \textit{mobility of care} trips, it is well-established that society undervalues this type of \emph{unpaid labor}. Public transit systems traditionally have reflected this bias against unpaid labor and \textit{mobility of care} by designing service and fares around traditional commuting patterns. Most systems are built to focus on providing more extensive and frequent service between residential neighborhoods and the downtown core, especially around peak commuting times prioritizing the travel needs of a commuter. In a resource-constrained environment where public transit agencies must make difficult decisions about where to cut or add service, change fares, and build infrastructure, such biases, albeit unintentional, can have profound impacts on those who rely on transit for \textit{mobility of care}. By shedding light on the scale of \textit{mobility of care} tripmaking, as well as the higher likelihood that such trips are performed by women, this research provides WMATA and other transit agencies with the knowledge and tools to measure, mitigate, and reverse these longstanding biases. We identified a handful of bus routes that serve a disproportionate number of \textit{mobility of care} trips and destinations. Service planning involving these routes (e.g. open stroller policy, service changes) should consider the \textit{mobility of care} and gender implications. Additionally, we identified that WMATA's current transfer policy offers built-in support \textit{mobility of care} trip-making patterns by allowing unlimited bus-to-bus transfers with the same SmarTrip card within two hours and offering fare discounts for rail-to-bus or bus-to-rail connections.

\begin{figure}[!h]
\centering
\includegraphics[width= .55\textwidth]{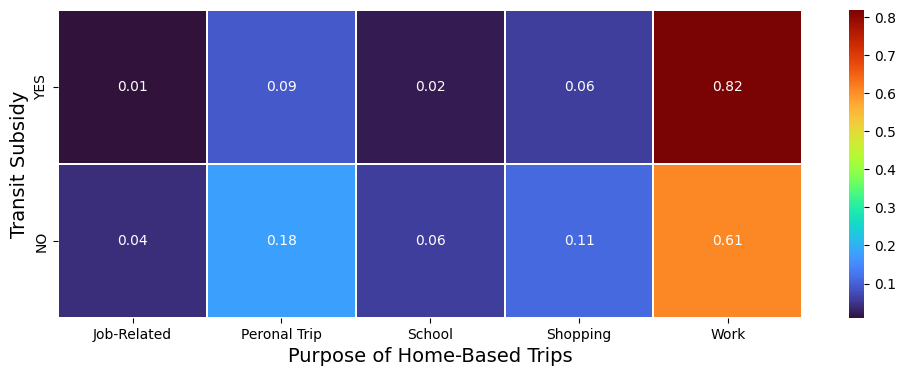}
\caption{\label{fig:metrobus_survey} Self-reported trip purpose (in percentage per trip purpose) for Metrobus customers identifying as women with and without the SmartBenefits subsidy.}
\end{figure}

Further research is necessary to capture a wider array of individuals who might not be covered in the existing subset of users registering their SmarTrip cards. As the literature suggests, income and employment status are strong variables affecting \textit{mobility of care} travel patterns. Despite the evident differences observed in gendered travel patterns, a limitation of this study is that it utilizes data that is most likely more reflective of the travel behavior of full-time employed individuals, who registered their cards to enroll in the SmartBenefits and other employer-subsidized programs. In responses analyzed from a WMATA customer survey for their bus system users, women respondents who reported having a transit subsidy indicated that 82\% of their journeys are work-related. This percentage drops to 61\% for women without a transit subsidy, who are currently far less likely to register their SmarTrip card. This shows the immense need for encouraging and incentivizing representative customer card data that would enable more robust gender-based analyses.

\section{Conclusions}

In this study, we propose and evaluate a data-driven workflow to identify and assess \textit{mobility of care} tripmaking behavior. We present a case study in the National Capital Region using data from the Washington Metropolitan Area Transit Authority (WMATA). Using ODX data, a name-based gender inference approach to associate a likely gender with fare cardholder name, and geospatial analysis, we observed that women-classified cards indeed travel and trip-chain more often to \textit{mobility of care} locations than men-classified cards. This deviation from gender parity is particularly observed during weekday mornings and afternoons, at times associated with higher activities at \textit{mobility of care} POIs. For a gender-balanced dataset, we observed that approximately 60\%-65\% of the trips to and from stops near daycare centers were made by women-classified cards. This trend was similarly observed at schools and grocery stores and found consistent for those places of interest that fall outside the city center. Additionally, we found women significantly more likely to exhibit accompanying patterns, consistently traveling to accompany those who may be children, elderly, or people with disabilities. The findings of this study validate the findings of other studies in the growing body of literature concluding that women are more likely to use public transit to conduct mobility \textit{mobility of care} trips compared to men.

Although this analysis focused on pre-COVID data, its implications are relevant post-COVID. As of September 2023, transit ridership in the United States remains depressed compared to pre-pandemic levels. At WMATA, ridership is still only around 60\% of pre-pandemic levels, for example. Whereas before COVID, some may have considered \textit{mobility of care} only as a matter of equity, there is no doubt now that transit agencies need to treat all trip purposes seriously as a matter of financial sustainability. The backbone of public transit in the United States prioritizes the traditional commuter, and traditional commuting patterns have been upended by the pandemic. 

\section{Acknowledgements}
The authors would like to thank WMATA for their generous support of this project, and the MIT SuperCloud and Lincoln Laboratory Supercomputing Center for providing high-performance computing resources that have contributed to the research results reported in this paper.

\section{Author Contribution Statement}
The authors confirm their contribution to the paper as follows: study conception and design: A.A., D.S., A.S, K.C., M.P.; data collection: D.S., A.A.; analysis and interpretation of results: D.S., A.A., A.S., K.C., M.P, G.P.; draft manuscript preparation: A.A., D.S., A.S, K.C., M.P., G.P., I.S., J.Z. All authors reviewed the results and approved the final version of the manuscript. %The authors do not have any conflicts of interest to declare.

\section{Declarations of Interest}
None.

%\nolinenumbers
\bibliography{references}

\clearpage
\appendix
\setcounter{figure}{0}
\setcounter{table}{0}
\section{Validating Name-based Gender Inference \label{appendix:a}}
\vspace{20pt}

\begin{figure}[!h]
\begin{subfigure}{.49\textwidth}
  \centering
  \includegraphics[width= \textwidth]{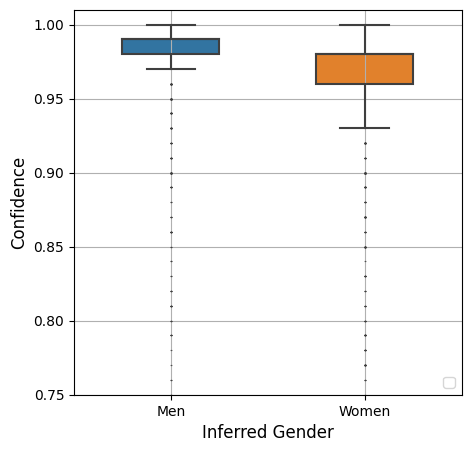}
  \caption{\label{fig:gender_probs} Overall.}
\end{subfigure}
\begin{subfigure}{.49\textwidth}
  \centering
  \includegraphics[width= \textwidth]{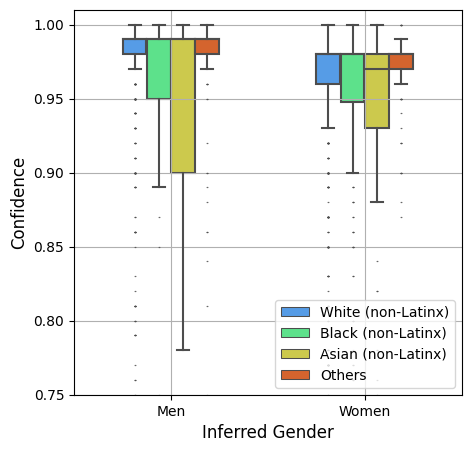}
  \caption{\label{fig:race_probs} By Race.}
\end{subfigure}
\caption{\label{fig:gender_race_probs} Confidence of name-based gender inference by the self-reported race of Metrobus survey respondents.}
\end{figure}

To assess the accuracy of our name-based gender inference pipeline, we conducted a comparison looking at self-reported gender information from one of WMATA's customer research team's surveys. The survey was conducted by reaching out to registered SmarTrip card holders and had $\sim$5,000 respondents, but when accounting for users with multiple SmarTrip cards, the dataset contains about 4,000 distinct Metro riders. Overall, we were able to accurately predict the self-reported gender of 94\% of the customers on the survey using first name inference. We did observe, however, that the inference confidence for women's names is slightly lower than that of the self-identifying men. When visualizing the confidence distribution of gender prediction across the self-reported races of the survey respondents, we find that lower gender confidence is associated with women overall, and more generally for Black and Asian names where the accuracy drops to an average of 87\%. Figure \ref{fig:gender_race_probs} shows the distributions of confidence across inferred gender and self-reported race, and Table \ref{tab:misclassifications} shows the raw error rate (percent of misclassification) by race.

While higher \emph{classification accuracy} is attainable by utilizing higher \emph{inference confidence cut-offs} (i.e. considering only cards where the confidence of the first-name-based gender inference pipeline is higher than 90\%), this analysis concludes that doing so would disproportionately impact women and riders of color, hence we keep the default cut-off point at 51\% (i.e. the binary gender with higher confidence is taken as the inferred gender for a given name).

\begin{table}[!h]
  \centering
  \renewcommand{\arraystretch}{1.75}
  \caption{Gender Inference Error Rate By Self-Reported Race For Metrobus Survey Respondents}
    \begin{tabular}{ccccc}
    \toprule
    \multicolumn{1}{l}{\textbf{Self-Reported Gender}} & \textbf{Race} & \textbf{Misclassifications} & \textbf{Total Count} & \textbf{Error (\%)} \\
    \hline
    \multirow{4}[8]{*}{\textbf{Women}} & White (non-Latinx) & 95    & 1,289 & 7.37 \\
\cline{2-5}          & Black (non-Latinx) & 35    & 332   & \textbf{10.54} \\
\cline{2-5}          & Asian (non-Latinx) & 19    & 130   & \textbf{14.62} \\
\cline{2-5}          & Other & 10    & 191   & 5.24 \\
    \hline
    \multirow{4}[7]{*}{\textbf{Men}} & White (non-Latinx) & 40    & 1,341 & 2.98 \\
\cline{2-5}          & Black (non-Latinx) & 10    & 152   & 6.58 \\
\cline{2-5}          & Asian (non-Latinx) & 7     & 110   & 6.36 \\
\cline{2-5}          & Other & 7     & 157   & 4.46 \\
    \end{tabular}%
  \label{tab:misclassifications}%
\end{table}

\clearpage
\setcounter{figure}{0}
\setcounter{table}{0}
\section{Sensitivity Analysis of Buffer Size For Identifying Metrobus Stops Serving Mobility of Care Places of Interest \label{appendix:b}}
\vspace{20pt}

As detailed in Algorithm 1 in the Methodology section, we used a 400m radius to evaluate stops and identify the nearest stop to a \textit{mobility of care} POI for every GTFS route pattern. We employ this radius to limit candidate stops and reduce the search space for each mobility of care POI. The WMATA system has approximately 11,000 bus stops, hence an exhaustive search for all the stops for every \textit{mobility of care} POI (or the POIs of any other use case of interest) would be quite inefficient. The FHWA guidelines suggest that an access/egress walking distance of ¼ - to - ½ mile (400m – 800m) is generally found acceptable by transit riders. Considering that the transit riders we’re interested in are presumed to accompany children, elderly, or carrying grocery bags, we use the lower threshold of 400m as an upper limit and evaluate only the stops within that radius to find the nearest GTFS-route-pattern stop for every \textit{mobility of care} location that falls within that \emph{buffer}.

\begin{figure}[!h]
\begin{subfigure}{.49\textwidth}
  \centering
  \includegraphics[width= \textwidth]{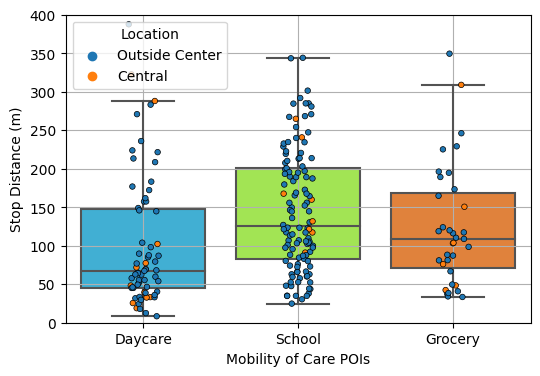}
  \caption{\label{fig:nearest_stop} Nearest Stop.}
\end{subfigure}
\begin{subfigure}{.49\textwidth}
  \centering
  \includegraphics[width= \textwidth]{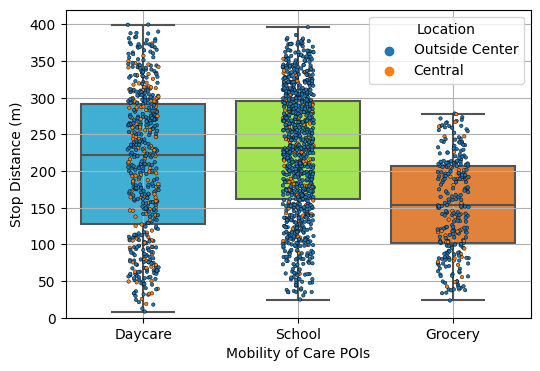}
  \caption{\label{fig:gtfs_stops} Nearest GTFS-Pattern Stop.}
\end{subfigure}
\caption{\label{fig:nearest_stops} Distribution of nearest stops to Mobility of Care POIs.}
\end{figure}

This assumes that the riders know and will only get off at the closest stop for each GTFS route pattern even if multiple stops fall within the 400m radius. For the \textit{mobility of care} POI stops in Washington D.C. identified via this method, we found the average POI to nearest bus stop distance to be \textbf{105 meters} (\textbf{74} meters for daycares, \textbf{125m} for schools, and \textbf{108m} for grocery stores). Figure \ref{fig:nearest_stops} shows the distribution of those stops, with Figure \ref{fig:nearest_stop} illustrating the absolute nearest bus stop to each POI, and Figure \ref{fig:gtfs_stops} showing the distribution of the nearest stop for all GTFS-route pattern that falls within that 400m radius to a mobility of care POI.

\newpage
Additionally, we looked at the impact of reducing the maximum distance of the nearest stop search from 400m to 200, 100, and 50 meters. Table 1 of this appendix shows the total number of \textit{mobility of care} POIs GTFS-pattern-stops (at 400m buffer) and the percentage that falls within each reduced radius. We see that a reasonable number of stops is preserved at each level of radius reduction, with the exception of schools where only 9\% of the stops believed to serve the schools can be found within 50 meters. This is attributable to the schoolyards and open spaces that separate many school buildings (where the centroid of the POI's geolocation would be) from the nearest main road with transit service. We found the results for gender ratio deviation to be consistent across all buffer sizes, albeit smaller buffers resulting in slightly more variability due to having fewer observations. Overall, out of 260 \textit{mobility of care} POIs we considered,  only 24 (9\%) had no Metrobus stop within 400 meters.

\begin{table}[!h]
\centering
\renewcommand{\arraystretch}{1.75}
\caption{Mobility of Care Stops Within Varying Maximum Distances From POIs}\label{tab:buffer}
\begin{tabular}{lccccc}\toprule
&{$\#$ \textbf{Stops} \par} &{\textbf{Within 200m} \par} &{\textbf{Within 100m} \par} &{\textbf{Within 50m} \par} \\\cmidrule{2-5}
\textbf{Daycares} &594 &85\% &68\% &33\% \\
\textbf{Schools} &1,029 &75\% &34\% &9\%  \\
\textbf{Grocery} &288 &86\% &43\% &23\%  \\
\bottomrule
\end{tabular}
\end{table}

\newpage
\setcounter{figure}{0}
\setcounter{table}{0}    
\section{Deviation from Gender Parity \label{appendix:c}}

\begin{figure}[!h]
\centering
\begin{subfigure}{.49\textwidth}
\includegraphics[width= \textwidth]{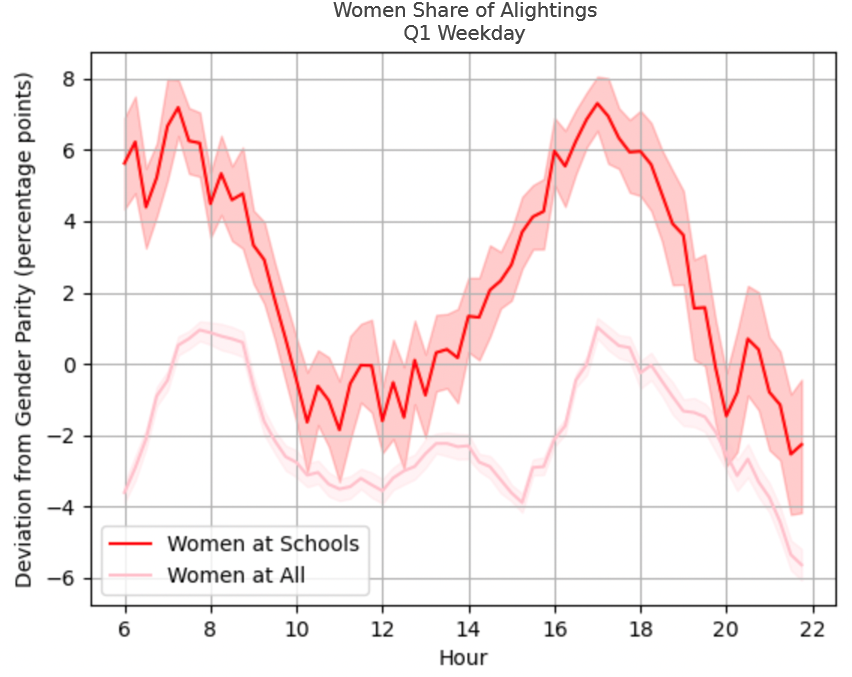}
\caption{\label{fig:schools_ratio_end}  Schools.}
\end{subfigure}
\begin{subfigure}{.49\textwidth}
\centering
\includegraphics[width= \textwidth]{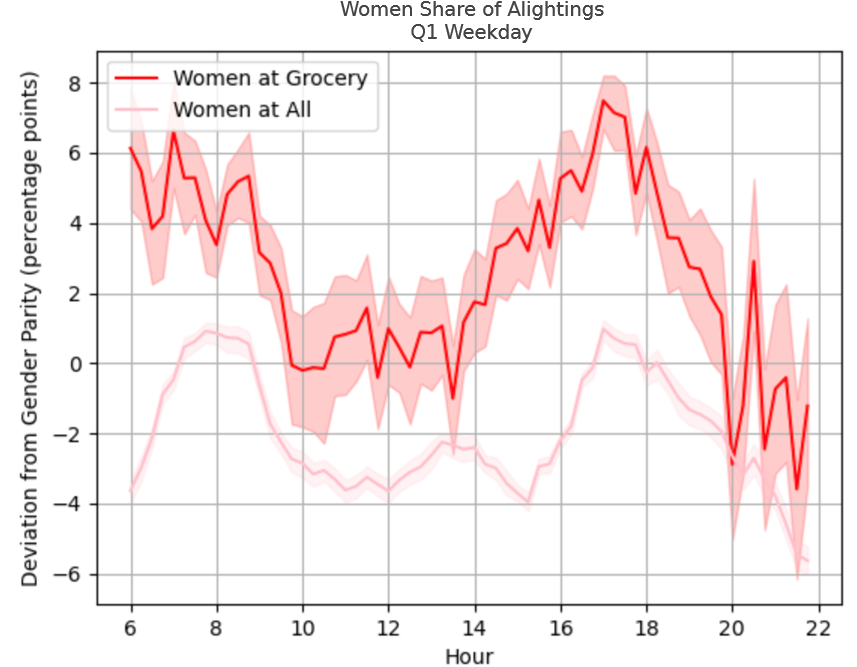}
\caption{\label{fig:grocery_end} Grocery stores.}
\end{subfigure}
\caption{\label{fig:school_grocery_endstops} Distribution of alightings at nearest stops to Mobility of Care POIs.}
\end{figure}

\begin{figure}[!h]
\begin{subfigure}{.33\textwidth}
  \centering
  \includegraphics[width= \textwidth]{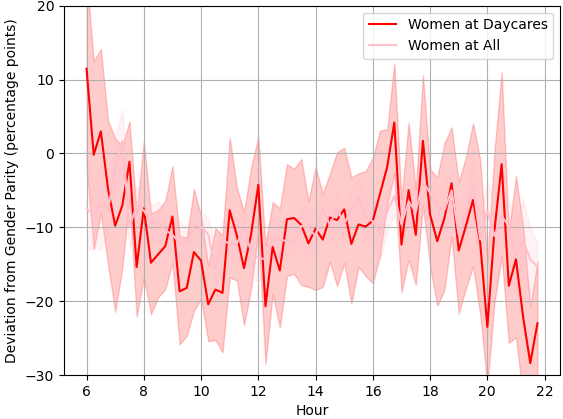}
  \caption{\label{fig:daycare_weekend} Daycares.}
\end{subfigure}
\begin{subfigure}{.33\textwidth}
  \centering
  \includegraphics[width= \textwidth]{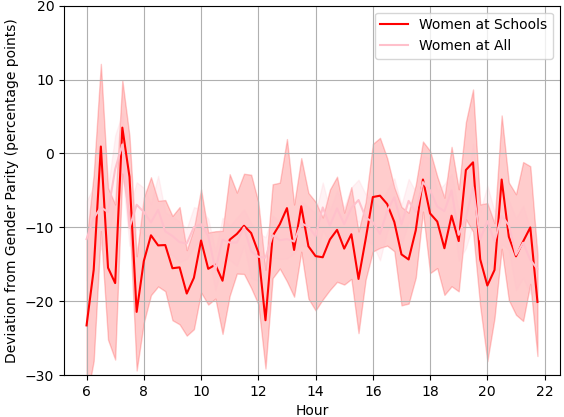}
  \caption{\label{fig:school_weekend} Schools.}
\end{subfigure}
\begin{subfigure}{.33\textwidth}
  \centering
  \includegraphics[width= \textwidth]{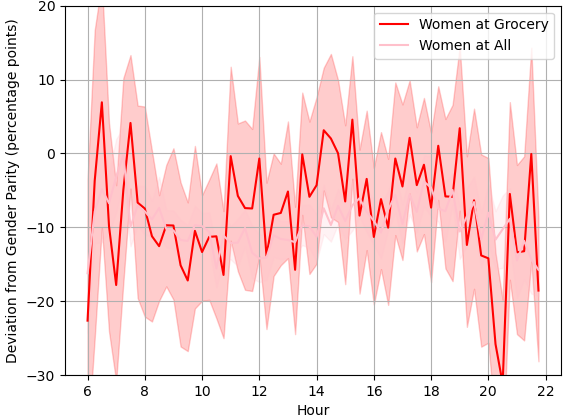}
  \caption{\label{fig:grocery_weekend} Grocery stores.}
\end{subfigure}
\caption{\label{fig:weekend} Gender ratio deviation from 50\% women at POIs during weekends.}
\end{figure}

\subsection{Statistical Relationship Between \textit{Mobility of Care} (MoC) and Gender}

\begin{figure}[!h]
    \centering
    \includegraphics[width=1\linewidth]{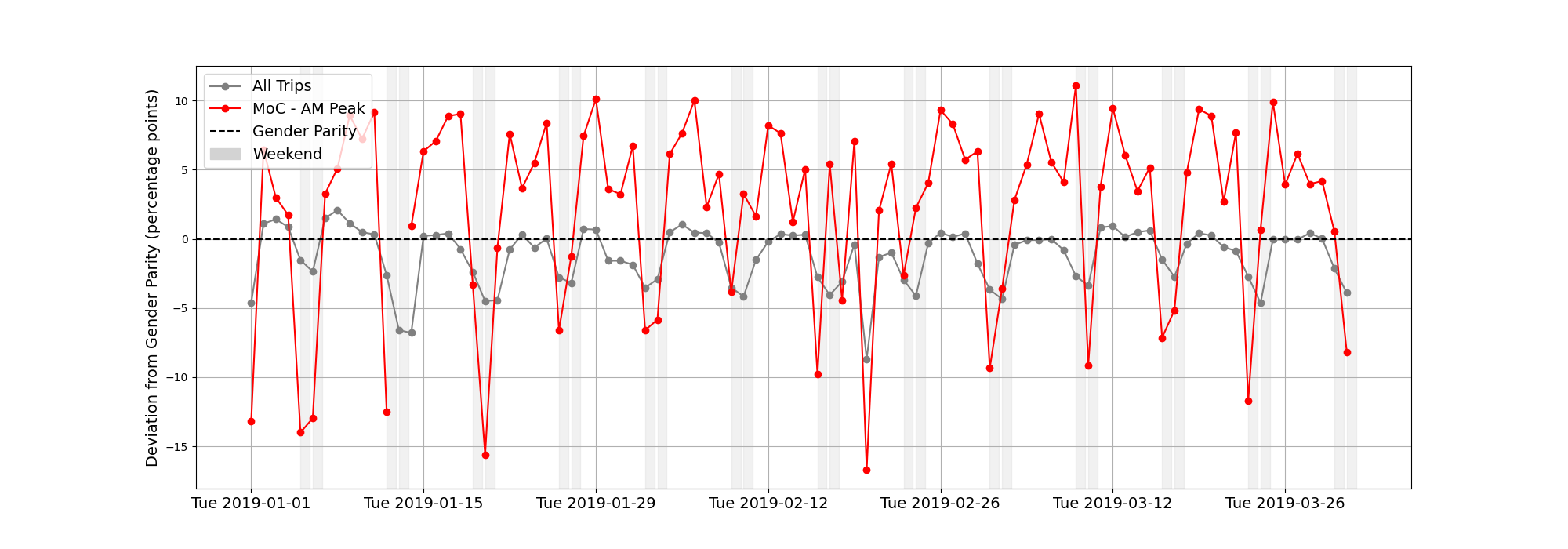}
    \caption{Deviation from gender parity for women for MoC trips compared to all trips, across Q1 2019}
    \label{fig:dev_gender_q1_2019_schools_daycares}
\end{figure}

Figure \ref{fig:dev_gender_q1_2019_schools_daycares} demonstrates the significant temporal relationship between MoC trips and gender. During weekdays, as illustrated in the figure, the share of women performing MoC trips is higher than that of men and higher than the share of women conducting all other (non-MoC) trips. On weekends (shaded in grey), the percentage of women traveling to MoC stops is comparable to the percentage of women traveling across the entire network with no significance. A few exceptions on weekdays are explainable by closed schools or holidays. For example, the deviation from gender parity on Monday, January 21, 2019, where approximately -4\% for both MoC trips and for all trips, was Martin Luther King Jr. Day, a Federal Holiday in the United States. Further, D.C schools were closed on Wednesday, February 20th, which is reflected in a large drop in women traveling to all stops and specifically to MoC stops. 

We also this relationship in terms of accompanying trips and the percentage of women conducting those trips across quarter 1 of 2019. Some exceptions, also observed in the previous figure, are during  Wednesday, February 20th, a holiday for D.C. schools and Martin Luther King Jr. Day, Monday, January 21st, which is also reflected in the uncharacteristically low rate of female travel for a Monday. This day is a federal holiday in the United States.

\begin{figure}[!h]
    \centering
    \includegraphics[width=1\linewidth]{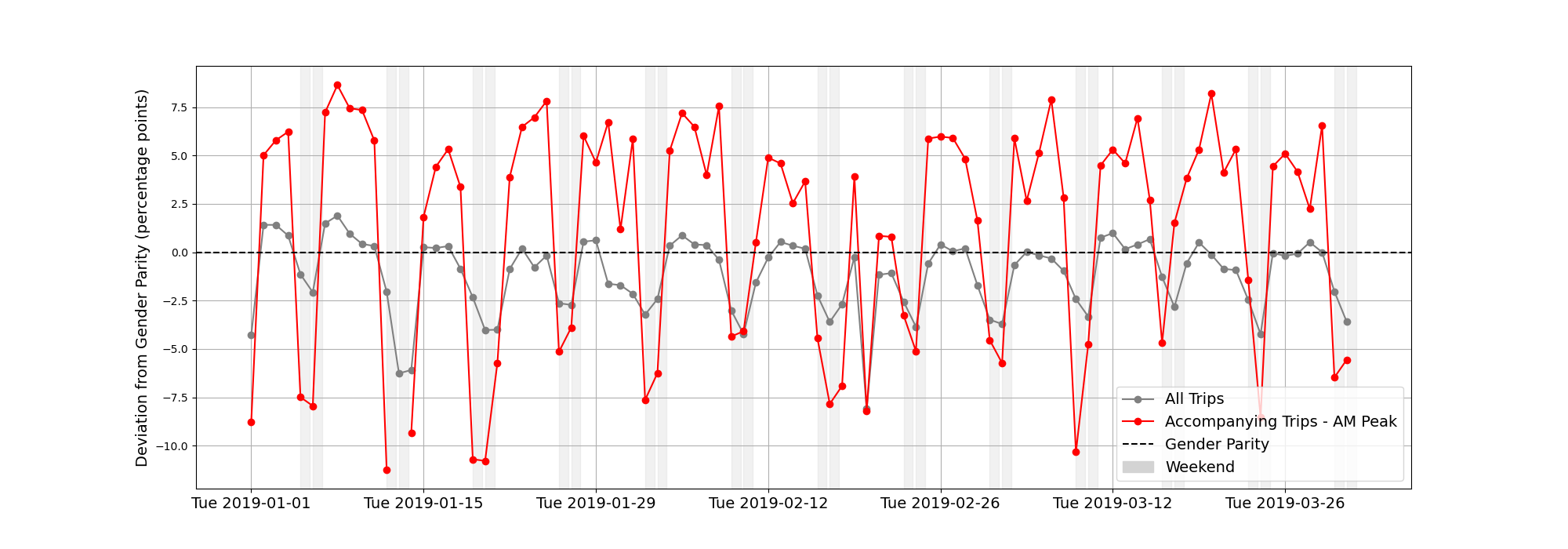}
    \caption{Deviation from gender parity for women for accompaniment trips compared to all trips, across Q1 2019}
    \label{fig:dev_gender_q1_2019_accompaniment}
\end{figure}

A Chi-Square Test of Independence was employed to determine whether MoC trips and the trips conducted by women are related to each other. The null hypothesis $(H_0)$ is that the proportion of trips that are MoC trips are the same for both men and women. This would suggest there is no relationship between MoC Trips and a SmarTrip card belonging to a woman. The alternative hypothesis $(H_A)$ is the proportion of trips that are MoC trips differ between genders. There are three conditions to apply a Chi-Square Test of Independence: 
\begin{enumerate}
    \item \textit{The two variables being tested are two categorical variables}. Indeed, both gender (in this study) and MoC are binary categorical variables. 
    \item \textit{The sample was randomly selected from the population}. When running the test, we took a random sample of 1 million trips from the full approximately 7 million trip dataset in Q1 of 2019.
    \item \textit{A minimum of 5 observations are expected in each group.} This is covered by the extensive dataset used for this study.
    \end{enumerate}

The Chi-Square tests reveal statistical significance between MoC trips and trips conducted by SmarTrip card belonging to women with a p-value less than 0.05. 

\begin{table}[!h]
    \centering
    \renewcommand{\arraystretch}{1.5}
    \begin{tabular}{lr}
        \toprule
        \textbf{Statistic} & \textbf{Value} \\
        \midrule
        Chi-Square Statistic & 168.220 \\
        p-value & $2.961 \times 10^{-37}$ \\
        Degrees of Freedom & 2 \\
        \bottomrule
    \end{tabular}
    \caption{Chi-Square Test Results: Gender v. MoC}
    \label{tab:chi_square}
\end{table}

\subsection{MoC vs. Non-MoC Trip Convenience Statistical Validation}
In order to evaluate whether a MoC trip exhibits different characteristics than non-MoC trips significantly, a Welch's T-Test was applied. A Welch's T-test can be applied to evaluate whether two populations have the same mean, without assuming they have equal variances. The null hypothesis is that MoC trips and non-MoC trips \emph{between the same origin-destination (O-D) pairs} should have the same in-vehicle time, and the same number of transfers (after accounting for the additional transfer by definition in MoC trips). 

The following three assumptions need to be verified in order to apply Welch's T-Test: \textbf{independence}, \textbf{normality}, and \textbf{no extreme outliers}. 

\begin{enumerate}
    \item \textbf{Independence}: Each row in the dataset represents a trip. The data contains multiple rows performed by across the same O-D pair. Thus, the power of the test may be inflated for both MoC and non-MoC trips if certain O-Ds are more dominant than others. A mixed-effects regression accounting for dependencies between observations across each O-D pair is later presented.
    \item \textbf{Normality}: An assumption in Welch's T-test is that the sub-populations are Normally distributed around the mean. We can observe the deviation from a normal distribution using a quantile-quantile plot. Both distributions for MoC and non-MoC travel times follow a normal distribution generally.
    \item \textbf{Outliers}: Extreme or influential outliers may influence the results of the test. The travel speed (euclidean distance divided by trip time) was within a reasonable range (0-25 mph). This suggests there are no outliers on the right tail-end of the distribution. Approximately 0.01\% of trips had in-vehicle times of more than 3 hours. Given the sparsity of the outliers in the dataset, it is unlikely they influenced the results of the test.
\end{enumerate}
\begin{table}[htbp]
  \centering
  \caption{Welch's T-Test: Convenience Metrics v. MoC Trip Across O-D Pairs}
  \begin{tabular}{|l|c|c|c|c|c|}
    \hline 
    \textbf{Variable} & \textbf{MoC} & \textbf{Non-MoC} & \textbf{Difference (MoC - Non-MoC)} & \textbf{T-Stat} & \textbf{P-Value}\\
    \hline
    In-Vehicle Time & 38.115 & 27.972 & 10.143 & 43.623 & $0.0^{***}$\\
    Transfers & 1.325 & 0.372 & 0.953 & 179.644 & $0.0^{***}$\\
    % Travel Speed & 8.82 & 19.76 & -10.94 & -359.59 & $0.0^{***}$\\
    \hline
  \end{tabular}
  \label{tab:ttest_convenience_metrics}
\end{table}

The results of the t-test (Table \ref{tab:ttest_convenience_metrics}) indicate a statistically significant difference in the means between MoC and non-MoC trips. Particularly, a MoC trip is on average 10 minutes longer than a non-MoC trip between the same O-D pair. The p-values are all below .05. The t-statistics are sufficiently high, however, because of the large number of data points (>10,000). The t-values, thus, are inflated and don't necessarily represent the power of the test. 

One way to reduce the inflated power of the t-test is to account for the effects of the O-D pairs. Using a mixed-effects regression, we can include each unique O-D pair as a variable. We can express this with the following equation. 

\begin{equation}
\text{in\_vehicle\_time}_{ij} = \beta_0 + \beta_1 \cdot \text{moc\_flag}_{ij} + u_i + \epsilon_{ij}
\end{equation}

where:
\begin{itemize}
    \item \(\text{in\_vehicle\_time}_{ij}\) is the dependent variable for the \( j \)-th observation in the \( i \)-th group.
    \item \(\beta_0 = 27.940\) is the fixed intercept (average trip time on the network).
    \item \(\beta_1 = 10.108\) is the fixed coefficient for the predictor \(\text{moc\_trip}_{ij}\).
    \item \(u_i \sim N(0, 514.486)\) is the random intercept for the \( i \)-th O-D pair.
    \item \(\epsilon_{ij} \sim N(0, 185.3478)\) is the residual error term for the \( j \)-th observation amongst the \( i \)-th O-D pair.
\end{itemize}

\begin{table}[htbp]
\centering
\caption{Mixed Linear Model: In-Vehicle Time Comparison between MoC and non-MoC Trips Across O-D Pairs}
\begin{tabular}{lccccc}
\hline
\textbf{In-Vehicle Time} & Coef. & Std.Err. & z & P$>|z|$ & [0.025 0.975] \\
\hline
Intercept & 27.940 & 0.098 & 286.032 & 0.000 & (27.748, 28.131) \\
moc\_trip & 10.108 & 0.250 & 40.354 & 0.000 & (9.617, 10.599) \\
\hline
\end{tabular}
\label{tab:mixed_convenience}
\end{table}

The results of this test (Table \ref{tab:mixed_convenience}) demonstrate that even when we account for the dependence assumption, the differences in means between non-MoC and MoC trips are statistically significant. MoC trips take on average $\sim$10 minutes longer in-vehicle than Non-MoC trips between the same O-D pair. 

\newpage
\setcounter{figure}{0}
\setcounter{table}{0}    
\section{Accompaniment Fare Products \label{appendix:d}}

\begin{figure}[!h]
\begin{subfigure}[t]{0.50\textwidth}
  \centering
  \includegraphics[width=\textwidth, height=2.8in, valign=t]{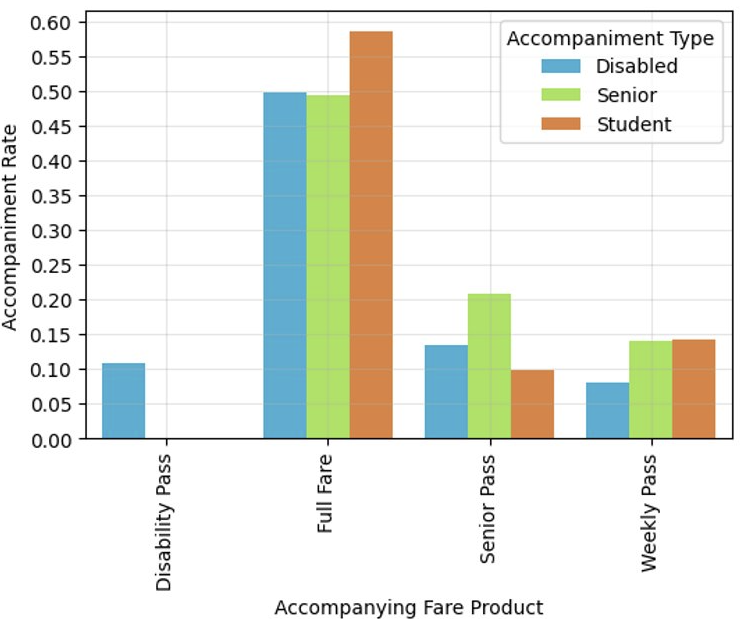}
  \vspace{13pt}
  \caption{Low-rate accompaniers (1 accompaniment per week).}
  \label{fig:acc_product_low}
\end{subfigure}
\begin{subfigure}[t]{0.50\textwidth}
  \centering
  \includegraphics[width=\textwidth, height=3.0in, valign=t]{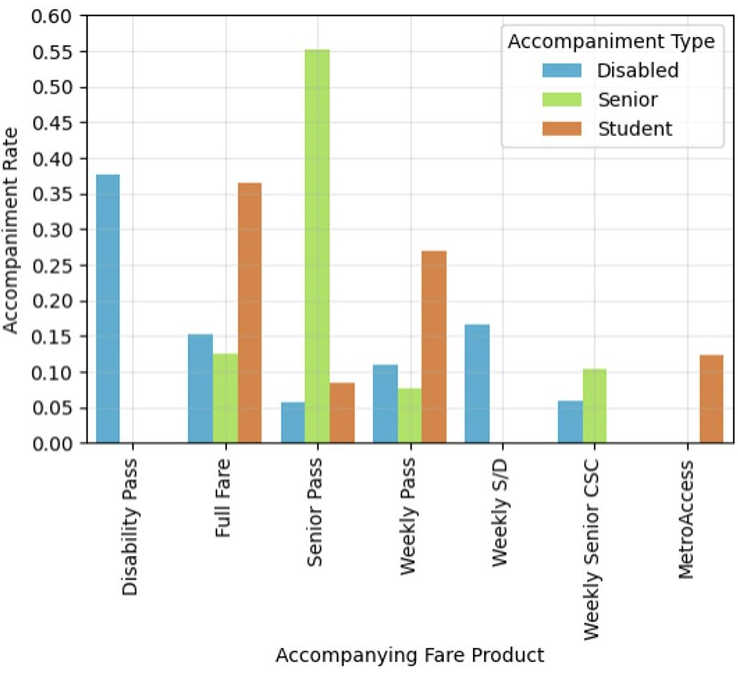}
  \caption{High-rate accompaniers (3+ accompaniments per week).}
  \label{fig:acc_product_high}
\end{subfigure}
\caption{Percentage of accompaniment types conducted by different WMATA fare products.}
\label{fig:acc_fare_distribution}
\end{figure}

The ratio of accompaniment trips of the three types we analyzed (disabled, senior, and students) conducted by accompanying cards using various WMATA fare products is illustrated below. For brevity, we only show products that were used in 3\% or more of accompaniments of a given type. It can be observed that high-rate accompaniers switched from relying predominantly on the pay-as-you-go full-fare product, which was used for 50+\% of accompaniments of all kinds for low-rate accompaniers, to more specific fare products that better serve their purpose. It can be observed that for cards with high accompaniment rates seniors and individuals with disabilities tend to travel together, this is clearly observed in Figure \ref{fig:acc_product_high} where 35\% of all disability accompaniments are with another disability pass holder and nearly 55\% of recurrent accompaniment for seniors being with another card with a senior fare product. A subset of individuals accompanying individuals with a disability are included in the disability pass category. We have not yet investigated whether this is strictly a behavior of individuals using those fare products recurrently traveling together or if it is attributable to traveling in groups with coordinated care services.

An interesting finding from this analysis is that the accompaniment rate for students using the full-fare product remains elevated, even for individuals with a high accompaniment rate, where 35\% of accompanying trips are conducted paying in full-fare. It is worth noting that for schools within Washington D.C., the Metrobus system is the de facto school bus system and all K-12 students receive the District's \emph{Kids Ride Free} fare product. While the share of student accompaniers using a weekly pass significantly increases, this might indicate a gap in fare offerings for this particular group of caregivers compared to the existing fare products for individuals accompanying seniors and individuals with disabilities.

\end{document}